\documentclass[11pt]{article}

\usepackage{epsfig,graphics}
\usepackage{amssymb,amsmath,amsthm,bm}

\newtheorem{theo}{Theorem}
\newtheorem{lem}{Lemma}

\newtheorem{cor}{Corollary}
\newtheorem{mrem}{Remark}

\newtheorem{example}{Example}

\def\R{{\mathbb R}}

\def\RR{{\mathbb R}}

\def\ga{\alpha}
\def\gga{\gamma}

\def\go{\omega}

\def\gb{\beta}

\def\gs{\sigma}
\def\gl{\lambda}
\def\wt{\widetilde}
\def\n{\noindent}

\def\gp{{\prime}}

\def\ep{\epsilon}
\def\vep{\varepsilon}

\def\gt{\triangle}

\def\b0{{\bf 0}}
\def\1{{\bf 1}}

\def\bd{{\bf d}}

\def\cC{\mathcal C}
\def\cE{\mathcal E}

\def\cS{\mathcal {S}}

\def\rem{{\rm rem}}
\def\Rem{{\rm Rem}}

\def\err{{\rm res}}
\def\Err{{\rm Res}}
\def\bd{{\rm bd}}
\def\Bd{{\rm Bd}}

\def\pd{{\partial}}

\def\wh{\widehat}

\def\wt{\widetilde}

\begin{document}

\title{Analysis of Adaptive Synchrosqueezing Transform with a Time-varying Parameter\thanks{This work was supported in part by the National Natural Science Foundation of China 
under grants 61373087, 11871348, 61872429
and Simons Foundation under grant 353185.}
}

\author{Jian Lu${}^{1}$,  Qingtang Jiang${}^2$,  and Lin Li${}^3$
}

\date{}

\maketitle


\bigskip
{\small 1. Shenzhen Key Laboratory of Advanced Machine Learning and Applications, College of Mathematics and Statistics, Shenzhen University, Shenzhen 518060, P.R. China. E-mail: jianlu@szu.edu.cn}

{\small 2. Department of Mathematics and Statistics, 
University of Missouri-St. Louis, St. Louis,  MO 63121, USA.  E-mail: jiangq@umsl.edu}

{\small 3. School of Electronic Engineering, Xidian University, Xi${}'$an,  710071, P.R. China.  

E-mail: lilin@xidian.edu.cn}

\begin{abstract}

The synchrosqueezing transform (SST) was developed recently to separate the components of  non-stationary multicomponent signals. The continuous wavelet transform-based SST (WSST) reassigns the scale variable of the continuous wavelet transform of a signal to the frequency variable and sharpens the time-frequency representation. The WSST with a time-varying parameter, called the adaptive WSST, was introduced very recently  in the paper  ``Adaptive synchrosqueezing transform with a time-varying parameter for non-stationary signal separation". The well-separated conditions of non-stationary multicomponent signals with the adaptive WSST and a method to select the time-varying parameter were proposed in that paper. In addition,  simulation experiments in that paper show that the adaptive WSST is very promising in estimating the instantaneous frequency of a multicomponent signal, and in accurate component recovery. However the theoretical analysis of the adaptive WSST has not been studied. In this paper, we carry out such analysis and obtain error bounds for 
component recovery with the adaptive WSST and the 2nd-order adaptive WSST. These results provide a mathematical guarantee to non-stationary multicomponent signal separation with the adaptive WSST. 
\end{abstract}

\bigskip 

\bigskip 

\centerline{{\bf AMS Mathematics Subject Classification:} 42C40,  42C15, 42A38}

\section{Introduction}

Most real signals such as EEG and bearing signals are non-stationary multicomponent signals given by
\begin{equation}
\label{AHM0}
x(t)=A_0(t)+\sum_{k=1}^K x_k(t), \qquad x_k(t)=A_k(t) e^{i 2\pi \phi_k(t)},
\end{equation}
with $A_k(t), \phi_k'(t)>0$, where  $A_0(t)$ is the trend, and $A_k(t), 1\le k\le K$, are called the instantaneous amplitudes  and $\phi'_k(t)$ the instantaneous frequencies.  Modeling a non-stationary signal  $x(t)$ as in \eqref{AHM0}
is important to extract information hidden in $x(t)$.
 The empirical mode decomposition (EMD) algorithm along with the Hilbert spectrum analysis (HSA)
is a popular method to decompose and analyze nonstationary signals \cite{Huang98}.
EMD decomposes a nonstationary signal as a superposition of intrinsic mode functions (IMFs) and then the instantaneous frequency of each IMF is calculated by HSA which results in a representation of the signal as in \eqref{AHM0}. The properties of EMD have been studied and variants of EMD have been proposed to improve the performance in many articles, see e.g.  \cite{Cicone20, HM_Zhou16, HM_Zhou20, Flandrin04, LCJJ19, HM_Zhou09,Oberlin12a, Rilling08,van20,Y_Wang12,Wu_Huang09, Xu06}. 
A weakness of EMD is that it can easily lead to mode mixtures or artifacts, namely undesirable or false components \cite{Li_Ji09}. In addition, there is no mathematical theorem to guarantee the recovery of the components.

Recently  the continuous wavelet transform-based  synchrosqueezed transform (WSST) was developed in  \cite{Daub_Lu_Wu11} to separate the components of a non-stationary multicomponent signal. 
In addition, the short-time Fourier transform-based SST (FSST) was also proposed in \cite{Thakur_Wu11} and further studied in \cite{Wu_thesis, MOM14} for this purpose.
To provide sharp representations for signals with significant frequency changes, the  2nd-order FSST and  the 2nd-order WSST were introduced in \cite{MOM15} and \cite{OM17} respectively,
and the theoretical analysis of them was carried out in \cite{BMO18} and \cite{Pham17b} respectively.
Other SST related methods include the generalized WSST \cite{Li_Liang12},
a hybrid empirical mode decomposition-SST computational scheme \cite{Chui_Walt15},
the synchrosqueezed wave packet transform \cite{Yang15}, WSST with vanishing moment wavelets \cite{Chui_Lin_Wu15}, the demodulation-transform based SST \cite{Wang_etal14, Jiang_Suter17, WCSGTZ18},
higher-order FSST \cite{Pham17}, signal separation operator \cite{Chui_Mhaskar15} and empirical signal separation algorithm \cite{LCJJ19}. In addition, the synchrosqueezed curvelet transform for two-dimensional mode decomposition was introduced in \cite{YangY14} and the statistical analysis of synchrosqueezed transforms has been studied in \cite{Yang18}.

SST provides an alternative to the EMD method and its variants, and it overcomes some limitations of the EMD scheme \cite{Flandrin_Wu_etal_review13,Meig_Oberlin_M12}. SST has been used in multiple applications including machine fault diagnosis \cite{Li_Liang_fault12,WCSGTZ18},
crystal image analysis \cite{Yang_Crystal_18,Yang_Crystal_15},
welding crack acoustic emission signal analysis \cite{HLY18}, and medical data analysis  
\cite{Wu_heartbeat17,Wu_breathing14, Wu_sleep15}. 

Most of the  WSST and FSST algorithms available in the literature are based on a continuous wavelet or a window function with a  fixed window, which means high time resolution and frequency resolution cannot be obtained simultaneously.  Recently the R${\rm \acute e}$nyi entropy-based adaptive SST was proposed in \cite{Wu17} and the adaptive FSST with the window function containing time and frequency parameters was studied in \cite{Saito17}. Very recently the authors of \cite{CJLS18, LCHJJ18, LCJ18} considered 
the 2nd-order adaptive FSST and WSST with a time-varying parameter.  They obtained the well-separated condition for multicomponent signals  using linear frequency modulation signals to approximate a non-stationary signal at any local time. The experimental results  show that the adaptive SST is very promising in estimating the instantaneous frequency of  a multicomponent signal, and in accurate component recovery.
However the theoretical analysis of the adaptive SST has not been carried out.
The goal of this paper is to study the theoretical analysis of the adaptive WSST.
We obtain the error bounds for  component recovery with the adaptive WSST and the 2nd-order adaptive WSST. 

The rest of this paper is organized as follows. In Section 2,  we  briefly review  WSST, the 2nd-order WSST, the adaptive WSST and the 2nd-order adaptive WSST. We study the theoretical analysis of the (1st-order) adaptive WSST and that of  the 2nd-order adaptive WSST in Sections 3 and 4 respectively.  In both cases, we obtain the error bounds for component recovery. The proofs of  theorems and lemmas are presented in the appendices.

\section{Synchrosqueezed transform}

In this section we briefly review the continuous wavelet transform (CWT)-based synchrosqueezed transform (WSST) and the adaptive WSST.
A function $\psi(t) \in L_2(\R)$ is called a continuous wavelet (or an admissible wavelet) if it satisfies (see e.g. \cite{Dau_book, Meyer_book}) the admissible condition:
\begin{equation}
\label{def_C_psi}
0<C_\psi:=\int_{-\infty}^\infty |\wh \psi(\xi)|^2\frac {d\xi}{|\xi|}<\infty,
\end{equation}
where $\wh \psi$ is  the Fourier transform of $\psi(t)$ is defined by
\begin{equation*}
\wh \psi(\xi):=\int_{-\infty}^\infty \psi(t) e^{-i2\pi \xi t}dt.
\end{equation*}
The CWT of a signal $x(t)\in L_2(\R)$ with a continuous wavelet $\psi$ is defined by
\begin{equation}
\label{def_CWT}
W_x(a, b):=\int_{-\infty}^\infty x(t) \frac 1{a}\overline{\psi \Big(\frac{t-b}a\Big)} dt.
\end{equation}
The variables $a$ and $b$ are called the scale and time variables respectively. The signal $x(t)$ can be recovered by the inverse wavelet transform (see e.g. \cite{Chui_book,Chui_Jiang_book,Dau_book,Meyer_book})
$$
x(t)=\frac 1{C_\psi}\int_{-\infty}^\infty\int_{-\infty}^\infty W_x(a, b) \psi_{a, b}(t)db\;  \frac {da}{|a|}.
$$

A function $x(t)$ is called an analytic signal if it satisfies $\wh x(\xi)=0$ for $\xi<0$.
For an analytic continuous wavelets,  an analytic signal $x(t)\in L_2(\R)$ 
can be recovered by (refer to \cite{Daub_Maes96, Daub_Lu_Wu11}):
\begin{equation}
\label{CWT_recover_analytic_a_only}
x(b)=\frac 1{c_\psi} \int_0^\infty W_x(a, b) \frac {d a}a,
\end{equation}
where $c_\psi$ is defined by 
\begin{equation}
\label{def_c_psi}
0\not= c_\psi:=\int_0^\infty \overline{\wh \psi(\xi)} \frac {d\xi}{\xi}<\infty.
\end{equation}
Furthermore,
a real signal $x(t)\in L_2(\R)$ can be recovered by the following formula (see \cite{Daub_Lu_Wu11}):
\begin{equation}
\label{CWT_recover_real_a_only}
x(b)={\rm Re }\Big(\frac 2{c_\psi} \int_0^\infty W_x(a, b) \frac {d a}a \Big).
\end{equation}

The ``bump wavelet'' $\psi_{\rm bump}(x)$ defined by
\begin{equation}
\label{def_bump}
\wh \psi_{\rm bump}(\xi):=\begin{cases}
 e^{1-\frac 1{1-\gs^2(\xi-\mu)^2}}, \; & \hbox{if $\xi\in ( \mu-\frac 1\gs, \mu+\frac1\gs)$}\\
 0, \; & \hbox{elsewhere}, 
 \end{cases}
\end{equation}
where  $\gs>0, \mu>0$ with $\gs u>1$;  
and the (scaled) Morlet wavelet $\psi_{\rm Mor}(x)$ defined by
\begin{equation}
\label{def_Morlet0}
\wh \psi_{\rm Mor}(\xi):=e^{-2\gs^2 \pi^2(\xi-\mu)^2}-e^{-2\gs^2 \pi^2 (\xi^2+\mu^2)},
\end{equation}
where $\gs>0, \mu>0$, are the commonly used continuous wavelets.

Note that the CWT given above can be applied to a slowly growing $x(t)$ if the wavelet function $\psi(t)$ is 
in the Schwarz class $\cS$, the set of all such $C^\infty(\R)$ functions $f(t)$ that $f(t)$ and all of its derivatives are rapidly decreasing.

\subsection{CWT-based synchrosqueezing transform}

To achieve a sharper time-frequency representation of a signal, the synchrosqueezed wavelet transform (WSST) reassigns the scale variable $a$ to the frequency variable. For a given signal $x(t)$, let $\go_x(a, b)$ be the {phase transformation} \cite{Daub_Lu_Wu11} (also called ``instantaneous frequency  information" in \cite{Thakur_Wu11})  defined by
\begin{equation}
\label{def_phase}
\go_x(a, b) :={\rm Re}\Big( \frac{\partial _b W_x(a, b)}{i2\pi W_x(a, b)}\Big), \quad \hbox{for $W_x(a, b)\not=0$}.
\end{equation}
WSST is to reassign the scale variable $a$ by transforming CWT $W_x(a, b)$ of $x(t)$ to a quantity, denoted by
$T_{x, \gga}^{\gl}(\xi, b)$, on the time-frequency plane:
\begin{equation}
\label{def_SST}
T_{x, \gga}^{\gl}(\xi, b):=\int_{|W_x(a, b)|>\gga}  W_x(a, b) \frac 1{\gl} h\Big(\frac{\xi-\go_x(a, b)}\gl\Big) \frac {da}a,
\end{equation}
where throughout this paper $\gga>0$, $h(t)$ is a compactly supported function with certain smoothness and $\int_\R h(t)dt=1$, and $\int_{|W_x(a, b)|>\gga}$ means the integral  $\int_{\{a>0: \; |W_x(a, b)|>\gga\}}$ with $a$ ranging over the set $\{a: \; |W_x(a, b)|>\gga \; \hbox{and $a>0$}\}$.

We consider multicomponent signals $x(t)$ of \eqref{AHM0} with the trend $A_0(t)$ being removed, namely,
\begin{equation}
  \label{AHM}
  x(t)=\sum_{k=1}^K x_k(t)=\sum_{k=1}^K A_k(t) e^{i2\pi\phi_k(t)}
  \end{equation}
 with $A_k(t), \phi_k'(t)>0$. In addition, we assume that $\phi_{k-1}'(t)<\phi_k'(t), t\in \RR$ for $2\le k\le K$.

For $\vep>0$ and $\gt>0$, let ${\cal B}_{\vep, \gt}$ denote the set of multicomponent signals of \eqref{AHM} satisfying    the following conditions:
\begin{eqnarray}
\label{cond_basic0}&&A_k(t)\in C^1(\R)\cap L_\infty(\R), \phi_k(t)\in C^2(\R), \\
\label{cond_basic}&& A_k(t)>0, \; \inf_{t\in \R} \phi_k'(b)>0, \; \sup_{t\in \R} \phi_k'(b)<\infty\\
 \label{cond_phi_2nd_der}
 && |A'_k(t)|\le \vep \phi_k'(t),  \;  |\phi''_k(t)|\le \vep\phi_k'(t), \; t\in \R, \;
 M''_k:=\sup_{t\in \R} |\phi''_k(t)|<\infty, \\
\label{freq_resolution}&&\frac{\phi_k'(t)-\phi'_{k-1}(t)}{\phi_k'(t)+\phi'_{k-1}(t)}
\ge \gt, \; 2\le k\le K, t\in \R.
\end{eqnarray}
The condition \eqref{freq_resolution} is called the well-separated condition with resolution $\gt$.  For $1\le k\le K$, let ${\cal Z}_k$ be the zone in the
scale-time  plane defined by
\begin{equation}
\label{def_Zk0}
{\cal Z}_k:=\{(a, b): \; |1 - a\phi_k'(b)|< \gt\}.
\end{equation}
Then the well-separated condition \eqref{freq_resolution} implies that ${\cal Z}_k, 1\le k\le K$ are not overlapping.

In practice, for a particular signal $x(t)$, its CWT $W_x(a,b)$ lies in a region of the scale-time plane:
$$
\{(a, b): \; a_1(b)\le a\le a_2(b), b\in \RR\}
$$ 
for some $0<a_1(b),  a_2(b) <\infty$. That is $W_x(a,b)$ is negligible for  $(a, b)$ outside this region. Throughout this paper we assume for each $b\in \RR$, the scale $a$ is in the interval:
\begin{equation}
\label{a_interval} 
a_1(b)\le a\le a_2(b). 
\end{equation}

\bigskip

{\bf Theorem A.}  \cite{Daub_Lu_Wu11} {\it Let $x(t)\in {\cal B}_{\vep, \gt}$ with $0<\gt<1$ and $\wt \vep=\vep^{1/3}$. Let $\psi$ be a continuous wavelet in $\cS$ 
with supp($\wh \psi)\subseteq [1-\gt, 1+\gt]$.
 If $\vep$ is small enough, then  the following statements hold.

  {\rm (a)}
 For $(a, b)$ satisfying  $|W_x(a, b)|>\wt \vep$, there exists a unique $k\in \{1, 2, \cdots, K\}$ such that $(a, b)\in {\cal Z}_k$.

{\rm (b)} Suppose $(a, b)$ satisfies $|W_x(a, b)|>\wt \vep$ and  $(a, b)\in {\cal Z}_k$. Then
\begin{equation*}
|\go_x(a, b)-\phi_k'(b)|<  \wt \vep.
\end{equation*}

{\rm (c)} For any $k\in \{1, \cdots, K\}$,
\begin{equation} 
\label{reconstr_1st0}
 \Big| \lim_{\gl\to 0} \frac1{c_\psi}\int_{|\xi-\phi_k'(b)|<\wt \vep} T_{x, \wt \vep}^{\gl}(\xi, b)d\xi -x_k(b) \Big|\le  C(b)\wt \vep, 
\end{equation}
 where $C(b)<\infty$ is independent of $\wt \vep$.   
   }

\subsection{Adaptive WSST with a time-varying parameter}
In this paper  we consider continuous wavelets of the form
\begin{equation}
\label{wavelet_general}
\psi_\gs(t):= \overline{\frac 1\gs g\Big(\frac t\gs\Big)} e^{i2\pi \mu t}, 
\end{equation}
where $\gs>0, \mu>0$,  $g\in \cS$. 
In this paper, we let $\mu$ be a fixed positive number, e.g., one may set $\mu=1$. Thus for the simplicity of notation, we drop $\mu$ in $\psi_\gs(t)$. 
The parameter $\gs$  in $\psi_\gs(t)$ is also called the window width in the time-domain of wavelet $\psi_\gs(b)$.
The CWT of $x(t)$ with a time-varying parameter considered in \cite{LCJ18} is defined by
\begin{eqnarray}
\nonumber \wt W_x(a, b)\hskip -0.6cm &&:= 
\int_{-\infty}^\infty x(t) \frac 1{a}\overline{\psi_{\gs(b)} \Big(\frac{t-b}a\Big)} dt\\
&&\label{def_CWT_para0}
=\int_{-\infty}^\infty x(t) \frac 1{a \gs(b)} g\Big(\frac{ t-b}{a\gs(b)}\Big)e^{-i2\pi \mu \frac{t-b}a} dt\\
&&\label{def_CWT_para}
=\int_{-\infty}^\infty x(b+at) \frac 1{\gs(b)} g\Big(\frac t{\gs(b)}\Big)e^{-i2\pi \mu t} dt, 
\end{eqnarray}
where $\gs=\gs(b)$ is a positive and differentiable function of $b$.
We call $\wt W_x(a, b)$ the adaptive CWT of $x(t)$ with $\psi_\gs$.
If \begin{equation*}
0<c_\psi(b):=\int_0^\infty \overline{\wh \psi_{\gs(b)}(\xi)}
\frac {d\xi}{\xi}
=\int_0^\infty \wh g(\gs(b)(\mu-\xi)\frac {d\xi}{\xi}<\infty, 
\end{equation*}
then the original signal $x(b)$ can be recovered from $\wt W_x(a, b)$ (see \cite{LCJ18}):
\begin{equation}
\label{CWT_para_recover_analytic_a_only}
x(b)=\frac 1{c_\psi(b)} \int_0^\infty \wt W_x(a, b) \frac {d a}a,
\end{equation}
for analytic $x(t)$. In addition, if $\psi_{\sigma}$ is analytic, then for a real-valued $x(t)$, we have
\begin{equation*}
x(b)={\rm Re }\Big(\frac 2{c_\psi(b)} \int_0^\infty \wt W_x(a, b) \frac {d a}a \Big).
\end{equation*}

\bigskip
The condition $\wh \psi_{\gs(b)} (0)=\overline{\wh g(\gs(b) \mu)}=0$ for $\psi_\gs$  is required for $c_\psi(b)<\infty$.
When $g$ is bandlimited, i.e., $\wh g$ is compactly supported, to say supp$(\wh g) \subset [-\ga, \ga]$ for some $\ga>0$, then $\wh g(\gs(b) \mu)=0$ as long as 
\begin{equation}
\label{cond_gs}
\gs(b)>\frac \ga\mu, \; b\in \RR.
\end{equation}
 
If $\wh g$ is not compactly supported, we consider the ``support'' of  $\wh g$ outside which $\wh g(\xi)\approx 0$. More precisely, for a given small positive threshold $\tau_0$, if 
$|\wh g(\xi)|\le \tau_0$
for $|\xi|\ge \ga$ for some $\ga>0$, then we say $\wh g(\xi)$ is {\it essentially supported} in $[-\ga, \ga]$.
When $|\wh g(\xi)|$ is even and decreasing for $\xi\ge 0$,  
then $\ga$ is obtained by solving
\begin{equation}
\label{def_ga_general}
|\wh g(\ga)|=\tau_0.
\end{equation}
For example, when  $g$ is the Gaussian function defined by 
\begin{equation}
\label{def_g}
g(t)=\frac 1{\sqrt {2\pi}} e^{-\frac {t^2}2}, 
\end{equation}
then, with $\wh g(\xi)=e^{-2\pi^2 \xi^2}$,  the corresponding $\ga$ is given by
\begin{equation}
\label{def_ga}
\ga=\frac 1{2\pi}\sqrt{2\ln (1/{\tau_0 })} .
\end{equation}

For a non-bandlimited $g$, $\wh \psi_\gs(0)=0$ is not satisfied,  and in this case a second term is added to \eqref{wavelet_general} such that the resulting 
  \begin{equation*}
\wt \psi_\gs(t)=\frac 1\gs \overline{g\Big(\frac t\gs\Big)} e^{i2\pi \mu t}- {c_\gs}\overline{g\Big(\frac t\gs\Big)},
\end{equation*}
satisfies $\wh {\wt \psi}_\gs(0)=0$ (and hence  $c_{\wt \psi}(b)<\infty$), where $c_\gs$ is independent of $t$. See, for example, Morlet's wavelet $\psi_{\rm Mor}$ given in \eqref{def_Morlet0}, where a second term is required to assure $\wh \psi_{\rm Mor}(0)=0$. 

In this paper we study the error bound for individual component recovery by the adaptive WSST 
analogous to \eqref{reconstr_1st0}, where $c_\psi$ should be replaced by $c_\psi(b)$ for the adaptive WSST. However, instead of using $c_\psi(b)$, we will use a modified function of $b$ defined by
 \begin{equation}
\label{def_c_psi_para_ga}
c^\ga_\psi(b):=\int_{\mu-\frac \ga{\gs(b)}}^{\mu+\frac \ga{\gs(b)}} \overline{\wh \psi_{\gs(b)}(\xi)}
\frac {d\xi}{\xi}
=\int_{\mu-\frac \ga{\gs(b)}}^{\mu+\frac \ga{\gs(b)}}\wh g\big(\gs(b)(\mu-\xi)\big)
\frac {d\xi}{\xi}.
\end{equation}
As in \cite{LCJ18}, in this paper we always assume \eqref{cond_gs} holds.
Due to the condition \eqref{cond_gs}, $c^\ga_\psi(b)<\infty $ whether $g$ is bandlimited or not.

In the following we assume  $g\in \cS$, $|\wh g(\xi)|$ is even and decreasing for $\xi\ge 0$ unless $\wh g(\xi)$ is compactly supported. If $\wh g$ is not compactly supported, then $\ga$ is defined by \eqref{def_ga_general}  for a given small $\tau_0>0$.

\bigskip

Next we recall the adaptive WSST introduced in \cite{LCJ18}. First we denote
\begin{equation}
\label{def_psi_j}
g_1(t):=tg(t), \; g_2(t):=t^2g(t), \; g_3(t):=tg\rq{}(t). 
\end{equation}
We use $\wt W^{g_j}_x(a, b)$ and $\wt W^{g'}_x(a, b)$ to denote the adaptive CWT defined by \eqref{def_CWT_para} with $g$ replaced by $g_j$ and $g'$ respectively, where $1\le j\le 3$.

\bigskip

For  $x(t)=A  e^{i2\pi \xi_0 t}$ with $\xi_0>0$, one can show that (see \cite{LCJ18})
if $\wt W_x(a, b)\not=0$, then
\begin{equation}
\label{def_phase_para_complex}
\go^{\rm adp, c}_x(a, b):=\frac {\frac{\partial}{\partial b} \wt W_x(a, b)}{i2\pi \wt W_x(a, b)}+\frac {\gs'(b)}{i2\pi \gs(b)}
+ \frac {\gs'(b)}{\gs(b)}\frac {\wt W^{g_3}_x(a, b)}{i2\pi \wt W_x(a, b)},
\end{equation}
is $\xi_0$, the instantaneous frequency of $x(t)$. Note that ``c\rq\rq{} in $\go^{\rm adp, c}_x(a, b)$ means the complex version of the phase transformation. 
Hence, for a general $x(t)$,  \cite{LCJ18} defines $\go^{\rm adp}_x(a, b)$:=
Re$(\go^{\rm adp, c}_x(a, b))$, 
the real part of $\go^{\rm adp, c}_x(a, b)$,  as the phase transformation of the adaptive WSST. 
Then the (1st-order) adaptive WSST, denoted by
 $T_{x, \gga}^{\rm adp, \gl}$, is defined by
\begin{equation}
\label{def_adpWSST}
T_{x, \gga}^{\rm adp, \gl}(\xi, b):=\int_{|\wt W_x(a, b)|>\gga} \wt W_x(a, b) \frac 1{\gl}h\Big(\frac{\xi-\go^{\rm adp}_x(a, b)}\gl\Big) \frac{da}a.
\end{equation}

The 2nd-order adaptive WSST was proposed in  \cite{LCJ18}. To introduce the corresponding phase transformation,  the authors of  \cite{LCJ18} considered linear frequency modulation  signal (also  called linear chirp signal)
\begin{equation}
\label{def_chip_At}
x(t)=A e^{i2\pi \phi(t)}=A e^{i2\pi (\xi_0 t +\frac 12 r t^2)}.
\end{equation}
It was shown in \cite{LCJ18} that $\go_{x}^{\rm 2adp, c}(a, b)$ defined below is $\xi_0+rb$, the instantaneous frequency of $x(b)$:
\begin{equation}
\label{def_transformation_2nd_complex}
\go_{x}^{\rm 2adp, c}(a, b):=\frac {\frac{\partial}{\partial b} \wt W_x(a, b)}{i2\pi \wt W_x(a, b)}
+ \frac {\gs\rq{}(b)}{i2\pi \gs(b)} \Big(1+ \frac {\wt W^{g_3}_x(a, b)}{\wt W_x(a, b)}\Big)
- a \;\frac{\wt W^{g_1}_x(a, b)}{i2\pi \wt W_x(a, b)} R_0(a, b),
\end{equation}
for $(a, b)$ satisfying $\frac{\partial}{\partial a}\Big( a\frac { \wt W^{g_1}_x(a, b)}{\wt W_x(a, b)}\Big)\not=0$ and $\wt W_x(a, b)\not=0$, where
\begin{equation}
\label{def_R0}
R_0(a, b):=\frac 1{\frac {\partial}{\partial a}\Big(a \frac {\wt W^{g_1}_x(a, b)}{\wt W_x(a, b)}\Big) }\Big\{\frac {\partial}{\partial a}\Big(\frac {\frac {\partial}{\partial b} \wt W_x(a, b)}{\wt W_x(a, b)}\Big)+ \frac {\gs\rq{}(b)}{\gs(b)}
\frac {\partial}{\partial a}\Big(\frac {\wt W^{g_3}_x(a, b)}{\wt W_x(a, b)}\Big)\Big\}.
\end{equation}
 Then $\go_{x}^{\rm 2adp}$:=Re($\go_{x}^{\rm 2adp, c}$), 
 the real part of $\go_{x}^{\rm 2adp, c}$, is the phase transformation for the 2nd-order adaptive WSST. 
 
Here we consider two types of the 2nd-order adaptive WSSTs: 
\begin{equation}
\label{def_adp2ndWSST}
T_{x, \gga_1, \gga_2}^{\rm 2adp, \gl}(\xi, b):=\int_{\big\{a: \; |\wt W_x(a, b)|>\gga_1, \; |{\partial_a}(a{\wt W^{g_1}_x(a, b)}/{\wt W_x(a, b)})| >\gga_2\big\}} \wt W_x(a, b) \frac 1{\gl}h\Big(\frac{\xi-\go^{\rm 2adp}_x(a, b)}\gl\Big) \frac{da}a, 
\end{equation}
and  \begin{equation}
\label{def_adp2ndWSST_3}
S_{x, \gga_1, \gga_2}^{\rm 2adp,\gl}(\xi, b):=\int_{|\wt W_x(a, b)|>\gga_1} \wt W_x(a, b) \frac 1{\gl}h\Big(\frac{\xi-\go^{\rm 2adp}_{x,\gga_2}(a, b)}\gl\Big) \frac{da}a, 
\end{equation}
where $\go^{\rm 2adp}_{x,\gga_2}$ is the real part of $\go^{\rm 2adp, c}_{x,\gga_2}$ defined by 
$$
\go_{x, \gga_2}^{\rm 2adp, c}(a, b):=\left\{
\begin{array}{l}
\hbox{\rm quantity in \eqref{def_transformation_2nd_complex},  \quad  if  $|\wt W_x(a, b)|\not=0$ and $\Big|\frac{\partial}{\partial a}\Big(a\frac {\wt W^{g_1}_x(a, b)}{\wt W_x(a, b)}\Big)\Big| $} >\gga_2, \\
\hbox{\rm quantity in \eqref{def_phase_para_complex},  \quad  if  $|\wt W_x(a, b)|\not=0$ and $\Big|\frac{\partial}{\partial a}\Big(a\frac {\wt W^{g_1}_x(a, b)}{\wt W_x(a, b)}\Big)\Big| $} \le \gga_2. 
\end{array}
\right.
$$
Note that $\go_{x, \gga_2}^{\rm 2adp, c}(a, b)$  is $\go_{x}^{\rm 2adp, c}(a, b)$ with 
$\frac{\partial}{\partial a}\Big(a\frac {\wt W^{g_1}_x(a, b)}{\wt W_x(a, b)}\Big)\not=0$ described by  threshold $\gga_2>0$.

\bigskip
If $\gs(b)\equiv \gs$, a constant,  then $\go^{\rm 2adp}_x(a, b)$ is reduced to $\go^{\rm 2nd}_x(a, b)$ given by
\begin{equation}
\label{2nd_phase}
\go^{\rm 2nd}_x(a, b)=\left\{
\begin{array}{l}
{\rm Re}\Big\{\frac {\frac{\partial}{\partial b} W_x(a, b)}{i2\pi W_x(a, b)}\Big\}
- a{\rm Re}\Big\{ \frac{W_x^{g_1}(a, b)}{i2\pi W_x(a, b)}
\frac 1{\frac{\partial}{\partial a}\Big( a\frac {W_x^{g_1}(a, b)}{W_x(a, b)}\Big) }\frac{\partial}{\partial a}\Big(\frac {\frac{\partial}{\partial b} W_x(a, b)}{W_x(a, b)}\Big)\Big\},\\
\hskip 3cm \hbox{if $\frac{\partial}{\partial a}\Big(a\frac {W_x^{g_1}(a, b)}{W_x(a, b)}\Big)\not=0$ and $W_x(a, b)\not=0;$}\\
\\
{\rm Re}\Big\{\frac {\frac{\partial}{\partial b} W_x(a, b)}{i2\pi W_x(a, b)}\Big\},
\hbox{if $\frac{\partial}{\partial a}\Big(a\frac {W_x^{g_1}(a, b)}{W_x(a, b)}\Big)=0$,  $W_x(a, b)\not=0$.}
\end{array}
\right.
\end{equation}
Then we define the conventional 2nd-order WSSTs as
\begin{eqnarray*}
&& T_{x, \gga_1,\gga_2}^{\rm 2nd, \gl}(\xi, b)
:=\int_{\{|W_x(a, b)|>\gga_1, \; |{\partial_a}(a{W^{g_1}_x(a, b)}/{W_x(a, b)})| >\gga_2\}}  W_x(a, b) \frac 1{\gl}h\Big(\frac{\xi-\go^{\rm 2nd}_x(a, b)}\gl\Big) \frac{da}a,\\
&&\\
&& S_{x, \gga_1, \gga_2}^{\rm 2nd,\gl}(\xi, b)
:=\int_{|W_x(a, b)|>\gga_1} W_x(a, b) \frac 1{\gl}h\Big(\frac{\xi-\go^{\rm 2nd}_{x,\gga_2}(a, b)}\gl\Big) \frac{da}a. 
\end{eqnarray*}
 The conventional 2nd-order WSST was first introduced in \cite{OM17}. The reader refers to  \cite{OM17} for different
 phase transformations $\go_x^{\rm 2nd}(a, b)$.

\section{Analysis of adaptive WSST}

We assume
\begin{equation}
\label{freq_resolution_adp}
d\rq{}:=\min_{k\in \{1, \cdots, K\}}\min_{t\in \R}\frac{\phi_k'(t)-\phi'_{k-1}(t)}{\phi_k'(t)+\phi'_{k-1}(t)}> 0.
\end{equation}
Thus $x(t)$ satisfies the well-separated condition  \eqref{freq_resolution} with resolution $=d\rq{}/2$.
However, the value $d\rq{}$ may be very small. In this case, we cannot apply Theorem A directly.
The reason is that to guarantee the results in Theorem A to hold, the continuous wavelet $\psi$ needs to satisfy supp$(\wh \psi)\subseteq [1-\frac {d\rq{}}2, 1+\frac {d\rq{}}2]$ or at least
 $|\wh \psi(\xi)|$ is small for $|\xi-1|\ge {d\rq{}}/2$. If $d\rq{}$ is quite small, then $\psi$ has a very good frequency resolution, which implies by the uncertainty principle that  $\psi$  has a very poor time resolution, or equivalently $\psi$ has a very large time duration,
which results in large errors in instantaneous frequency estimate. 
We use  the adaptive CWT to adjust the time-varying  window width $\gs(b)$ at certain local time $t$ where the frequencies of two components are close.

In this section we consider the case that each component $x_k(t)=A_k(t)e^{i2\pi \phi_k(t)}$ is approximated locally by a sinusoidal signal. Here we consider conditions: 
\begin{equation}
\label{condition1} |A_k'(t)|\le \vep_1, \; |\phi''_k(t)|\le \vep_2, \; t\in \R, \; 1\le k\le K,
\end{equation}
for some small positive numbers $\vep_1, \vep_2$.
Let $\cC_{\vep_1, \vep_2}$ denote the set of multicomponent signals of \eqref{AHM} satisfying \eqref{cond_basic0}, \eqref{cond_basic},  \eqref{freq_resolution_adp} and \eqref{condition1}.

Let $x(t)\in \cC_{\vep_1, \vep_2}$.   Write $x_k(b+a t)$ as
\begin{eqnarray*}
x_k(b+a t)\hskip -0.6cm &&=x_k(b)e^{i2\pi \phi_k'(b) at }+(A_k(b+a t)-A_k(b))e^{i2\pi \phi_k(b+a t)}\\
&&\qquad +x_k(b)e^{i2\pi \phi_k'(b) at }\big(
e^{i2\pi (\phi_k(b+a t)-\phi_k(b)-\phi_k'(b) at)}-1\big).
\end{eqnarray*}
Then the adaptive CWT  $\wt W_x(a, b)$ of $x(t)$ defined by \eqref{def_CWT_para} with $g$ can be expanded as 
\begin{eqnarray}
\nonumber
\wt W_x(a, b)\hskip -0.6cm &&=\sum_{k=1}^K \int_\R x_k(b+a t)\frac 1{\gs(b)}g\Big(\frac t{\gs(b)}\Big) e^{-i2\pi \mu t}dt\\
\nonumber
&&=\sum_{k=1}^K \int_\R x_k(b)e^{i2\pi \phi_k'(b) at}\frac 1{\gs(b)}g\Big(\frac t{\gs(b)}\Big) e^{-i2\pi \mu t}dt +\rem_0,
\end{eqnarray}
or
\begin{equation}
\label{CWT_approx_1st}
\wt W_x(a, b)
=\sum_{k=1}^K x_k(b)  \wh g\big(\gs(b)(\mu-a \phi_k'(b))\big) +\rem_0,
\end{equation}
where $\rem_0$ is the remainder for the expansion of  $\wt W_x(a, b)$ given by
\begin{eqnarray}
\label{def_rem0}
&&\rem_0:=\sum_{k=1}^K \int_\R \Big\{
(A_k(b+a t)-A_k(b))e^{i2\pi \phi_k(b+a t)}
\\
\nonumber  &&\qquad \qquad
+x_k(b)e^{i2\pi \phi_k'(b) at }\big(
e^{i2\pi (\phi_k(b+a t)-\phi_k(b)-\phi_k'(b) at)}-1\big)
\Big\}\frac 1{\gs(b)}g\Big(\frac t{\gs(b)}\Big) e^{-i2\pi \mu t}dt.
\end{eqnarray}
With $|A_k(b+a t)-A_k(b)|\le  \vep_1 a |t|$
and
$$
|e^{i2\pi (\phi_k(b+a t)-\phi_k(b)-\phi_k'(b) at)}-1|\le 2\pi |\phi_k(b+a t)-\phi_k(b)-\phi_k'(b) at|
\le
\pi\vep_2 a^2|t|^2,
$$
we have
\begin{eqnarray*}
|\rem_0|\hskip -0.6cm && \le \sum_{k=1}^K \int_\R  \vep_1 a | t| \frac 1{\gs(b)}|g\Big(\frac t{\gs(b)}\Big)| dt+ \sum_{k=1}^K A_k(b)\int_\R  \pi \vep_2 a^2 | t|^2 \frac 1{\gs(b)}|g\Big(\frac t{\gs(b)}\Big)|
 dt   \\
 &&=K \vep_1 I_1 a \gs(b)  +\pi  \vep_2 I_2 a^2 \gs^2(b) \sum_{k=1}^K A_k(b), 
\end{eqnarray*}
where 
\begin{eqnarray}
&&\label{def_In}
I_n:=\int_\R  | t^n g(t)| dt, \;  n=1, 2, \cdots,
\end{eqnarray}
Hence we have
\begin{equation}
\label{rem0_est}
|\rem_0|\le a \gs(b) \lambda_0(a, b),
\end{equation}
where
\begin{equation}
\label{def_lam0}
\lambda_0(a, b):=K \vep_1  I_1 +\pi  \vep_2 I_2 a \gs (b) \sum_{k=1}^K A_k(b). 
\end{equation}

Similarly $\wt W^{g'}_x(a, b)$ can be expanded as \eqref{CWT_approx_1st} with remainder  $\rem_0'$,
defined as $\rem_0$ in \eqref{def_rem0} with $g(t)$ replaced by $g'(t)$.
Then we have the estimate for $\rem_0'$ similar to \eqref{rem0_est}. More precisely, we have
\begin{equation}
\label{err_est}
|\rem_0'|\le a \gs(b) \wt \lambda_0(a, b),
\end{equation}
where
\begin{equation}
\label{def_tlam0}
\wt \lambda_0(a, b):=K \vep_1  \wt I_1   +\pi \vep_2 \wt I_2 a \gs(b) \sum_{k=1}^K A_k(b),
\end{equation}
with  
\begin{eqnarray}
&&
\label{def_tIn}
\wt I_n:=\int_\R  | t^n g'(t)| dt, \; n=1, 2, \cdots.
\end{eqnarray}

\begin{mrem}\label{rem:ConditionB}
 The condition \eqref{cond_phi_2nd_der}, which was considered in \cite{Daub_Lu_Wu11}, means that $A_k(t)$ and instantaneous frequency $\phi_k'(b)$ change slowly compared with $\phi_k(t)$. For $FSST$, \cite{MOM14} uses another condition for the change of   $A_k(t)$ and $\phi_k'(b)$:
\begin{eqnarray}
\label{cond_phi_2nd_der1}&& |A'_k(t)|\le \vep,  \;  |\phi''_k(t)|\le \vep, \; t\in \R.
\end{eqnarray}

Condition \eqref{condition1} is essentially the condition  \eqref{cond_phi_2nd_der1}. If $A_k(t), \phi_k(t)$ satisfy  \eqref{cond_phi_2nd_der}, then we have a similar error bound for the expansion of $\wt W_x(a, b)$. More precisely, suppose  $A_k(t), \phi_k(t)$ satisfy
\begin{equation}
\label{conditionB}
  |A'_k(t)|\le \vep_1 \phi_k'(t),  \;  |\phi''_k(t)|\le \vep_2\phi_k'(t), \; t\in \R, \;
 M''_k:=\sup_{t\in \R} |\phi''_k(t)|<\infty.
\end{equation}
Then {\rm (}see {\rm \cite{Daub_Lu_Wu11})}
\begin{eqnarray*}
&&|A_k(b+a t)-A_k(b)|\le  \vep_1 a|t| ( \phi_k'(b)+\frac 12 M''_ka |t|), \\
&&|\phi_k(b+a t)-\phi_k(b)-\phi_k'(b) at| \le \vep_2 a^2 t^2 ( \frac 12 \phi_k'(b)+\frac 16 M''_k a |t|).
\end{eqnarray*}
Thus, we can expand $\wt W_x(a, b)$ as \eqref{CWT_approx_1st} with $|\rem_0|\le a\gs(b) \lambda_0(a, b)$, where in this case  $\lambda_0(a, b)$ is
\begin{equation}
\label{CondtionB_est}
\lambda_0(a, b):=
\vep_1 \sum_{k=1}^K \big(I_1 \phi_k'(b)+\frac 12 M''_k I_2 a \gs(b)\big)
   +\pi  \vep_2 a \gs(b) \sum_{k=1}^K A_k(b)\big(I_2\phi_k'(b)+\frac 13 M''_k I_3a \gs(b)\big).
\end{equation}
With the condition of \eqref{conditionB}, we have an estimate $a\gs(b) \wt \lambda_0(a, b)$ for $\rem_0\rq{}$,
the remainder for the expansion of $\wt W_x^{g'}(a, b)$, where in this case
$\wt \lambda_0(a, b)$ is defined by \eqref{CondtionB_est} with $I_j$ replaced by $\wt I_j$.
In this paper we consider the condition \eqref{condition1}.
The statements for the theoretical analysis of the adaptive WSST with condition \eqref{conditionB} instead of  \eqref{condition1}
 are still valid as long as $\lambda_0(a, b)$ in \eqref{def_lam0},  $\wt \lambda_0(a, b)$ in
\eqref{def_tlam0},  $\Lambda_k(b)$ and $\Lambda\rq{}_k(b)$
{\rm (}below in \eqref{def_Lam0} and \eqref{def_tLam0} respectively{\rm )} and so on are replaced respectively by
that in \eqref{CondtionB_est} and
 similar terms. This also applies to the 2nd-order adaptive WSST in Section 4, where we will not repeat this discussion on the condition like \eqref{conditionB}.
\hfill $\blacksquare$
\end{mrem}

If the remainder $\rem_0$ in \eqref{CWT_approx_1st} is small, then
the term $x_k(b)\wh g\big(\gs(b)(\mu-a \phi_k'(b))\big)$  in \eqref{CWT_approx_1st} determines the scale-time zone of the adaptive CWT $\wt W_{x_k}(a, b)$ of the $k$th component $x_k(t)$ of $x(t)$. More precisely, if $g$ is band-limited,  to say supp($\wh g)\subset [-\ga, \ga]$ for some $\ga>0$, then $x_k(b)\wh g\big(\gs(b)(\mu-a \phi_k'(b))\big)$ lies  within the zone of the scale-time plane:
 $$
Z_k:=\Big\{(a, b): |\mu - a \phi_k'(b)|< \frac {\ga}{\gs(b)}, b\in \R\Big\}.
 $$
The upper and lower boundaries of $Z_k$ are respectively
$$
\mu - a \phi_k'(b)= \frac {\ga}{\gs(b)}\; \hbox{and} \;  \mu - a \phi_k'(b)=-\frac {\ga}{\gs(b)}
$$
or equivalently
$$
a=(\mu + \frac {\ga}{\gs(b)})/ \phi_k'(b)\; \hbox{and} \;  a=(\mu - \frac {\ga}{\gs(b)})/ \phi_k'(b).
$$
Thus $Z_{k-1}$ and $Z_k$ do not overlap (with $Z_{k-1}$ lying above $Z_k$ in the scale-time plane) if
\begin{equation}
\label{separated_cond_1st0}
(\mu + \frac {\ga}{\gs(b)})/ \phi_k'(b) \le (\mu - \frac {\ga}{\gs(b)})/ \phi_{k-1}'(b),
\end{equation}
or equivalently
\begin{equation}
\label{separated_cond_1st}
\gs(b)\ge \frac {\ga}\mu  \frac { \phi_k'(b)+\phi_{k-1}'(b)}{ \phi_k'(b)-\phi_{k-1}'(b)}, \; b\in \RR.
\end{equation}
Therefore the multicomponent signal $x(t)$ is well-separated (that is $Z_k\cap Z_{\ell}=\emptyset, k\not=\ell$), provided that $\gs(b)$ satisfies \eqref{separated_cond_1st} for $k=2, \cdots, K$. 

Observe that our well-separated condition \eqref{separated_cond_1st} is different from that in \eqref{freq_resolution} considered in \cite{Daub_Lu_Wu11}.

When $\wh g$ is not compactly supported, let $\ga$ be the number defined by \eqref{def_ga_general}, namely assume  $\wh g(\xi)$ is {\it essentially supported} in $[-\ga, \ga]$.
Then $x_k(b)\wh g\big(\gs(b)(\mu-a \phi_k'(b))\big)$
 lies within the scale-time zone $Z_k$ defined by 
\begin{equation}
\label{def_Zk}
Z_k:=\Big\{(a, b): |\wh g\big(\gs(b)(\mu-a \phi_k'(b))\big) |>\tau_0, b\in \R\}=
\{(a, b): |\mu -a \phi_k'(b)|< \frac {\ga}{\gs(b)}, b\in \R\Big\}.
\end{equation}
Thus if the remainder $\rem_0$ in \eqref{CWT_approx_1st} is small,
$\wt W_{x_k}(a, b)$  lies within $Z_k$ and hence,
the multicomponent signal $x(t)$ is well-separated provided that  $\gs(b)$ satisfies \eqref{separated_cond_1st} for $2\le k\le K$. In this section we assume that \eqref{separated_cond_1st} with $k=2, \cdots, K$ holds for some $\gs(b)$.

From \eqref{rem0_est} and \eqref{err_est}, we have that
for $(a, b)\in Z_k$,
\begin{eqnarray}
&&\label{def_Lam0}
\frac {|\rem_0|}{a\gs(b)}\le \Lambda_k(b):=K \vep_1  I_1 +\pi  \vep_2 I_2
\frac{\mu \gs(b)+\ga}{\phi'_k(b)}  \sum_{j=1}^K A_j(b),\\
&&\label{def_tLam0}
\frac {|\rem'_0|}{a\gs(b)}\le \wt \Lambda_k(b):=K \vep_1  \wt I_1 +\pi  \vep_2 \wt I_2
\frac{\mu \gs(b)+\ga}{\phi'_k(b)}\sum_{j=1}^K A_j(b).
\end{eqnarray}

Here we remark that in practice $\phi'_k(t), 1\le k\le K$ are unknown.  However the condition in \eqref{freq_resolution} considered in the seminal paper \cite{Daub_Lu_Wu11} on SST and that in \eqref{separated_cond_1st} involve $\phi'_k(t)$. 
Like paper \cite{Daub_Lu_Wu11}, our paper establishes theoretical theorems which guarantee the recovery of components, namely, we provide conditions under which the components can be recovered. 

\bigskip 

Next we present our analysis results on the adaptive WSST in Theorem \ref{theo:main_1st} below,
where $\ga$ is defined by \eqref{def_ga_general}, and
throughout this paper,  $\sum_{\ell\not=k}$ denotes $\sum_{\ell\in \{1, \cdots, K\}\backslash\{k\}}$.
Recall that we assume that the scale variable $a$ lies in the interval \eqref{a_interval}. 
Throughout this section, we may assume that
\begin{equation}
\label{def_a2}
a_1=a_1(b):=\frac{\mu-\ga/\gs(b)}{\phi_K\rq{}(b)}\le a\le a_2=a_2(b):=
\frac{\mu+\ga/\gs(b)}{\phi_1\rq{}(b)}.
\end{equation}
In addition, we denote 
$$
\rho_{\ell, k}(b):=\begin{cases}
\gs(b)\mu -\big(\gs(b)\mu+\ga\big)\frac{\phi_\ell\rq{}(b)}{\phi_k\rq{}(b)}, & \hbox{if $\ell <k$},\\
\big(\gs(b)\mu-\ga\big)\frac{\phi_\ell\rq{}(b)}{\phi_k\rq{}(b)}-\gs(b)\mu, & \hbox{if $\ell >k$}.
\end{cases}
$$
Then one can obtain from \eqref{separated_cond_1st0} that for any $(a, b)\in Z_k$, the following inequality holds:
\begin{equation}
\label{rho_ineq}
|\gs(b)\big(\mu- a \phi_\ell\rq{}(b)\big)| >\rho_{\ell, k}(b). 
\end{equation}

\begin{theo}
\label{theo:main_1st} Suppose $x(t)\in \cC_{\vep_1, \vep_2}$ for some small $\vep_1, \vep_2>0$.
Let $Z_k$ be the scale-time zone defined by \eqref{def_Zk}.
Then we have the following.

{\rm (a)} Suppose $\wt \ep_1$ satisfies  $\wt \ep_1 \ge  a_2(b)\gs(b) \Lambda_1(b)+ \tau_0 \sum_{k=1}^K A_k(b)$, where $a_2(b)$ is given in \eqref{def_a2}.
Then for $(a, b)$ with $|\wt W_x(a, b)|>\wt \ep_1$,
there exists a unique $k\in \{1, 2, \cdots, K\}$ such that $(a, b)\in Z_k$.

{\rm (b)} For $(a, b)$ with $|\wt W_x(a, b)|\not=0$, we have
\begin{equation}
\label{transformation_approx}
\go_x^{\rm adp, c}(a, b)-\phi_k'(b)= \frac{\Rem_1}{i2\pi \wt W_x(a, b)},
\end{equation}
where  
$$ 
\Rem_1:=i2\pi \Big(\frac \mu a-\phi_k'(b)\Big)\rem_0 -  \frac{ \rem_0'}{a \gs(b)}+
i2\pi \sum_{\ell\not=k} x_\ell(b) \big(\phi'_\ell(b)-\phi_k'(b)\big)
\wh g\big(\gs(b)(\mu -a \phi_\ell'(b))\big).
 $$ 
Hence,  for $(a, b)$ satisfying $|\wt W_x(a, b)|>\wt \ep_1$ and  $(a, b)\in Z_k$, we have
\begin{equation}
\label{transformation_est}
|\go_x^{\rm adp}(a, b)-\phi_k'(b)|<  \bd_k,
\end{equation}
where
\begin{eqnarray}
\label{def_bd1}
\bd_k\hskip -0.6cm &&:=
\frac 1{\wt \ep_1}\big(\ga \Lambda_k(b)+\frac 1{2\pi}\wt \Lambda_k(b)\big)
+\frac 1{\wt \ep_1}\sum_{\ell\not=k} A_\ell(b)|\phi_\ell' (b)-\phi_k' (b)| \; \big|\wh g\big(\rho_{\ell, k}(b)\big)\big|. 
\end{eqnarray}

{\rm (c)} For a $k\in \{1, \cdots, K\}$, suppose that $\wt \ep_1$ satisfies the condition in part {\rm (a)}
and that $\bd_\ell$ in part {\rm (b)} satisfies 
$\max\limits_{1\le \ell \le K} \{\bd_\ell\}\le \frac 12 L_k(b)$, where 
\begin{equation}
\label{def_Lk}
L_k(b):=\min\{\phi'_k(b)-\phi'_{k-1}(b), \phi'_{k+1}(b)-\phi'_k(b)\}.
\end{equation}
Then for $\wt\ep_3$ satisfying
 $\max\limits_{1\le \ell \le K} \{\bd_\ell\}\le \wt \ep_3\le \frac 12 L_k(b)$, 
 we have
\begin{equation}
 \label{reconstr_1st}
 \Big| \lim_{\gl\to 0} \frac 1{c^\ga_\psi(b)}\int_{|\xi-\phi_k'(b)|<\wt \ep_3} T_{x, \wt \ep_1}^{\rm adp,  \gl}(\xi, b)d\xi -x_k(b) \Big|\le \wt{\bd}_k,
\end{equation}
where $c^\ga_\psi(b)$ is defined by \eqref{def_c_psi_para_ga}, and 
\begin{equation}
 \label{def_bd2}
\wt{\bd}_k:=\frac 1{|c^\ga_\psi(b)|} \Big\{
\wt \ep_1 \ln \frac {\mu \gs(b)+\ga}{\mu \gs(b)-\ga}  + \frac{2\ga}{\phi_k'(b)}
\Lambda_k(b)  +\sum_{\ell\not=k }A_\ell(b)  m_{\ell, k}(b)\Big\}
\end{equation}
with
$$
m_{\ell, k}(b):=\int_{\mu-\frac \ga{\gs(b)}}^{\mu+\frac \ga{\gs(b)}}\wh g\big(\gs(b)(\mu- \frac{\phi'_\ell(b)}{\phi'_k(b)} \xi)\big) \frac {d\xi}{\xi}.
$$
\end{theo}

\bigskip
The proof of Theorem \ref{theo:main_1st}(b) needs the following lemma.   
\begin{lem}
\label{lem:lem0} Let $\wt W_x(a, b)$ be the adaptive CWT of $x(t)$. Then
\begin{equation}
\label{result_partial_tV}
\pd_b\wt W_x(a, b)=\Big(\frac {i2\pi \mu}a-\frac {\gs'(b)}{\gs(b)}\Big)\wt W_x(a, b)
- \frac {\gs'(b)}{\gs(b)}\wt W^{g_3}_x(a, b)-\frac 1{a\gs(b)} \wt W^{g'}_x(a, b).
\end{equation}\end{lem}

We provide the proofs of Theorems \ref{theo:main_1st} and Lemma \ref{lem:lem0} in Appendices A and C respectively. In the rest of this section, we give some remarks on the results presented in Theorem \ref{theo:main_1st}.

\bigskip
\begin{mrem}
\label{rem:bandlimited}
When $\wh g(\xi)$ is supported in $[-\ga, \ga]$, then the condition in Theorem \ref{theo:main_1st}  part  {\rm (a)} for $\wt \ep_1$ is reduced to $\wt \ep_1 \ge a_2(b)\Lambda_1(b)$. Furthermore, in this case
$c^\ga_\psi(b)=c_\psi(b)$, and for $\ell\not= k$, 
$\wh g\big(\gs(b)(\mu- a \phi_\ell'(b))\big)=0$ for 
$(a, b)\in Z_k$ and $m_{\ell, k}(b)=0$.
Hence $\bd_k$ and $\wt \bd_k$ in  \eqref{def_bd1} and \eqref{def_bd2} are respectively
$$
\bd_k=\frac 1{\wt \ep_1}\big(\ga \Lambda_k(b)+\frac 1{2\pi}\wt \Lambda_k(b)\big), \quad
\wt\bd_k=\frac 1{|c_\psi(b)|} \Big(
\wt \ep_1 \ln \frac {\mu \gs(b)+\ga}{\mu \gs(b)-\ga}  + \frac{2\ga}{\phi_k'(b)}
\Lambda_k(b)\Big).
$$

Also,  Theorem \ref{theo:main_1st} can be written as Theorem A. To show this, in the following let us just consider the case $\vep_1=\vep_2$ for simplicity. Write
 $\Lambda_k(b), \wt \Lambda_k(b)$ defined by  \eqref{def_Lam0} and \eqref{def_tLam0} respectively as
 $$
 \Lambda_k(b)=\vep_1\gl_k(b), \; \wt \Lambda_k(b)=\vep_1\wt \gl_k(b),
$$
with
$$
 \gl_k(b):= K  I_1 +\pi   I_2 \frac{\mu \gs(b)+\ga}{\phi_k\rq{}(b)} \sum_{k=1}^K A_k(b), \;
 \wt \gl_k(b):= K \wt I_1 +\pi  \wt I_2  \frac{\mu \gs(b)+\ga}{\phi_k\rq{}(b)} \sum_{k=1}^K A_k(b).
 $$
 Let $\wt \ep_1=\vep_1^{1/3}$. If $\vep_1$ is small enough such that
 \begin{equation}
 \label{ep1_cond}
 \wt \ep_1\le \min\Big\{\frac 1{\sqrt{a_2(b)\gl_1(b)}},
 \frac 1{\max\limits_{1\le \ell \le K}\{\ga \lambda_\ell(b)+\frac 1{2\pi}\wt \lambda_\ell(b)\}}, \frac 12 
 L_k(b)\Big\},
 \end{equation}
  then $\wt \ep_1 \ge a_2(b) \wt \ep_1^3 \lambda_1(b)=  a_2 (b) \vep_1  \lambda_1(b)= a_2(b)\Lambda_1(b)$, and
  $$
  \max\limits_{1\le \ell \le K} \{\bd_\ell\}=\frac {\vep_1}{\wt \ep_1}\max\limits_{1\le \ell \le K} \big\{\ga \lambda_k(b)+\frac 1{2\pi}\wt \lambda_k(b)\big\}\le \frac {\vep_1}{\wt \ep_1} \frac 1{\wt \ep_1}=
  {\wt \ep_1}  \le \frac 12 L_k(b). 
  $$
  Thus the conditions in Theorem \ref{theo:main_1st} are satisfied, and the following corollary follows from
  Theorem \ref{theo:main_1st} immediately {\rm (}with $\wt \ep_3=\wt \ep_1${\rm )}.
 \end{mrem}

\begin{cor}
\label{cor:theo_standand_form}  Suppose $x(t)\in \cC_{\vep_1, \vep_1}$ for some small $\vep_1>0$, and  supp($\wh g) \subseteq [-\ga, \ga]$. Let $\wt \ep_1=\vep_1^{1/3}$. If $\vep_1$ is small enough such that \eqref{ep1_cond} holds, then we have the following.

  {\rm (a)}
 For $(a, b)$ satisfying  $|\wt W_x(a, b)|>\wt \ep_1$,  there exists a unique $k\in \{1, 2, \cdots, K\}$ such that $(a, b)\in Z_k$.

{\rm (b)} For $(a, b)$ satisfying $|\wt W_x(a, b)|>\wt \ep_1$ and  $(a, b)\in Z_k$, we have
\begin{equation*}
|\go_x^{\rm adp}(a, b)-\phi_k'(b)|<  \wt \ep_1.
\end{equation*}

{\rm (c)} For any $k$, $1\le k\le K$,
\begin{equation*}
 \Big| \lim_{\gl\to 0} \frac1{c_\psi(b)}   \int_{|\xi-\phi_k'(b)|<\wt \ep_1} T_{x, \wt \ep_1}^{\rm adp,  \gl}(\xi, b)d\xi -x_k(b) \Big|\le \frac 1{|c_\psi(b)|} \Big(
\wt \ep_1 \ln \frac {\mu \gs(b)+\ga}{\mu \gs(b)-\ga}   + {2\ga \wt \ep_1^3}\frac{\gl_k(b)}{\phi_k'(b)}\Big).
\end{equation*}
\hfill $\blacksquare$
\end{cor}

\begin{mrem}
 \label{rem:back_regular_WSST}
When $\gs(b)\equiv \gs$, a constant, $T_{x, \wt \ep_1}^{\rm adp,  \gl}(\xi, b)$ is the regular WSST
$T_{x, \wt \ep_1}^{\gl}(\xi, b)$ defined by \eqref{def_SST}. Suppose supp$(\wh g)\subseteq [-\ga, \ga]$.
 Then Corollary \ref{cor:theo_standand_form} is Theorem A with condition \eqref{condition1}.
\hfill $\blacksquare$
\end{mrem}

\begin{mrem}
\label{rem:theo_standand_form}
When $\wh g(\xi)$ is not supported on $[-\ga, \ga]$, but $|\wh g(\xi)|$  decays fast at $|\xi|\to \infty$, then the terms in the summation
$\sum_{\ell\not=k} $  for $\bd_k$ in \eqref{def_bd1} will be small as long as $\ga$ is quite large {\rm (}hence $\tau_0$ is very small{\rm )}. More precisely, 
from \eqref{separated_cond_1st0}, we have
$$
(\mu\gs(b)+\ga)\frac {\phi_{k-1}'(b)}{ \phi_k'(b)}\le \mu \gs(b)-\ga.
$$
Thus 
$$
 \rho_{k-1, k}(b)\ge \mu \gs(b)-\big(\mu\gs(b)-\ga\big)=\ga. 
$$
Similarly, we have $\rho_{k+1, k}(b)\ge \ga$. 
Recall that we assume that $|\wh g(\xi)|$ is decreasing on $\xi\ge 0$.  
Hence 
$$
|\wh g\big(\rho_{k\pm 1, k}(b)\big)|\le |\wh g\big(\pm \ga)|=\tau_0.
$$
The quantities $|\wh g\big(\rho_{\ell, k}(b) \big)|$ for other $\ell \not= k-1, k, k+1$ are smaller than $\tau_0$ also since $\rho_{\ell, k}(b)$ are  larger than $\ga$. 
As an example, let us consider the case when $g$ is the Gaussian function given in \eqref{def_g}. If we let $\ga=1$, then 
$$
\wh g(1)=2.675  \times 10^{-9}. 
$$
Thus even in practice $\wt \ep_1$ is small, for example $\wt \ep_1=10^{-4}$ or $10^{-5}$, and  hence  $1/\wt \ep_1$  is large, 
but the term in the summation $\sum_{\ell\not=k}$ for $\bd_k$ in \eqref{def_bd1} is still very small. 

\bigskip

For the functions $m_{\ell, k}(b)$ in \eqref{def_bd2}, we have
\begin{eqnarray*}
&&|m_{\ell, k}(b)|
\le \int_{\mu-\frac \ga{\gs(b)}}^{\mu+\frac \ga{\gs(b)}}\Big |\wh g\Big(\gs(b)\big(\mu- \frac{\phi'_\ell(b)}{\phi'_k(b)} \xi\big)\Big)\Big| \frac {d\xi}{\xi}\\
&&\le \int_{\mu-\frac \ga{\gs(b)}}^{\mu+\frac \ga{\gs(b)}}\tau_0 \frac {d\xi}{\xi}
=\tau_0  \ln \frac {\mu \gs(b)+\ga}{\mu \gs(b)-\ga} \approx \frac{2\ga }{\mu \gs(b)}\tau_0.
\end{eqnarray*}
 Thus $|m_{\ell, k}(b)|$ could be small if $\tau_0$ is small.
To summarize, in the case that $\wh g$ is not compactly supported, 
the statements in Corollary \ref{cor:theo_standand_form} still hold if the same conditions are satisfied and that  $\ga$ is large enough {\rm (}and hence $\tau_0$ is small enough{\rm )}.
\hfill $\blacksquare$
\end{mrem}

\begin{mrem} 
Observe that in Corollary \ref{cor:theo_standand_form}, $\wt \ep_1=\vep_1^{1/3}$. In \cite{Daub_Lu_Wu11} and \cite{MOM14} on theoretical analysis on WSST and FSST, $\wt \ep_1$ and $\vep_1$ have the same relationship. 
It means that if  $\wt \ep_1$ is small, then $\vep_1=\wt \ep_1^3 $ will be very small.  In other words, theoretically, 
to have small error bounds for the instantaneous frequency estimate, $|A^\gp_k(t)|$ and $|\phi^{\gp \gp}_k(t)|$ must be very small, which means $x(t)$ is essentailly a superposition of sinusoidal signals. This is the reason for that in practice WSST and FSST work well for sinusoidal signals, but not for signals with fast changing instantaneous frequency. The 2nd-order SSTs were introduced for signals with fast changing instantaneous frequency. We provide the analysis of 2nd-order adaptive WSST in the next section. 
\end{mrem}
Before moving on to the next section, we consider an example to show the recovery error bound $\wt {\bd}_k$ in 
\eqref{reconstr_1st}.

\begin{figure}[th]
	\centering
	\begin{tabular}{cc}
		\resizebox{3.0in}{2.0in}{\includegraphics{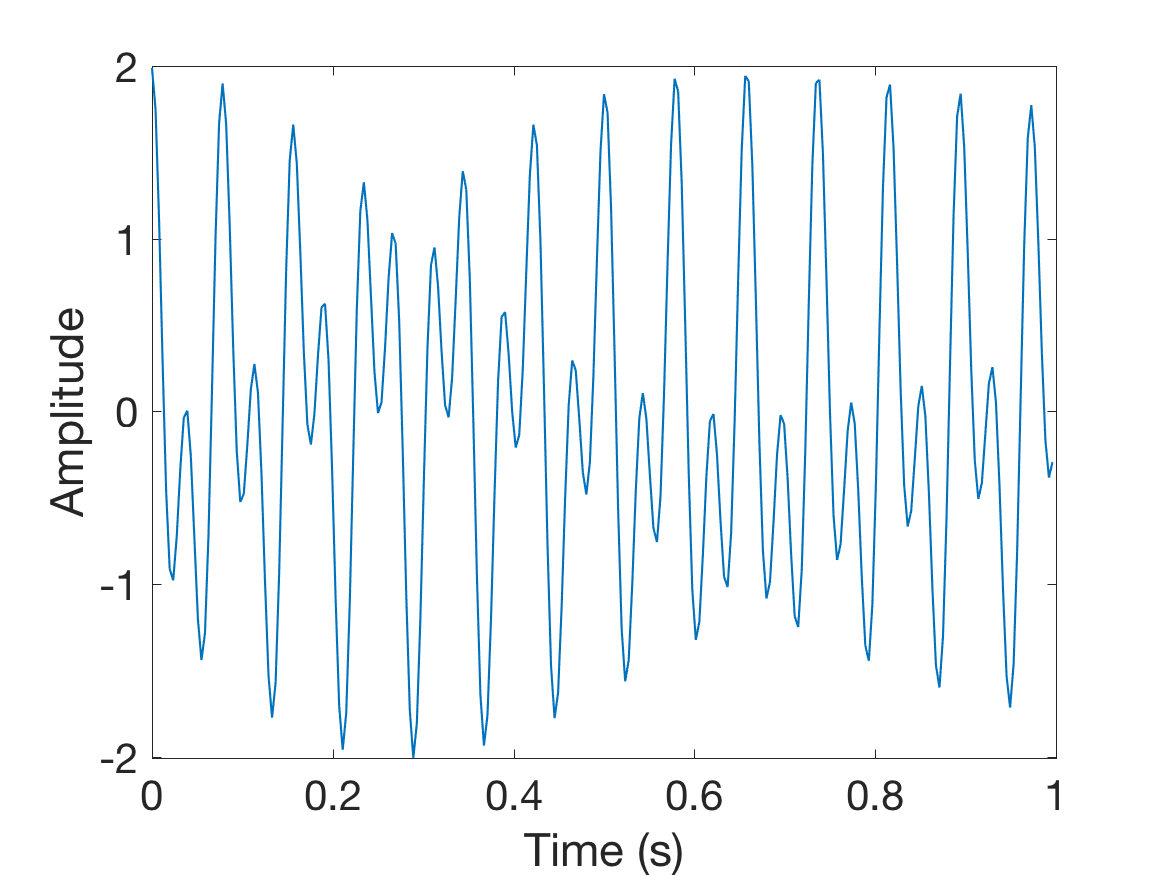}}\quad & \\
\resizebox{3.0in}{2.0in}{\includegraphics{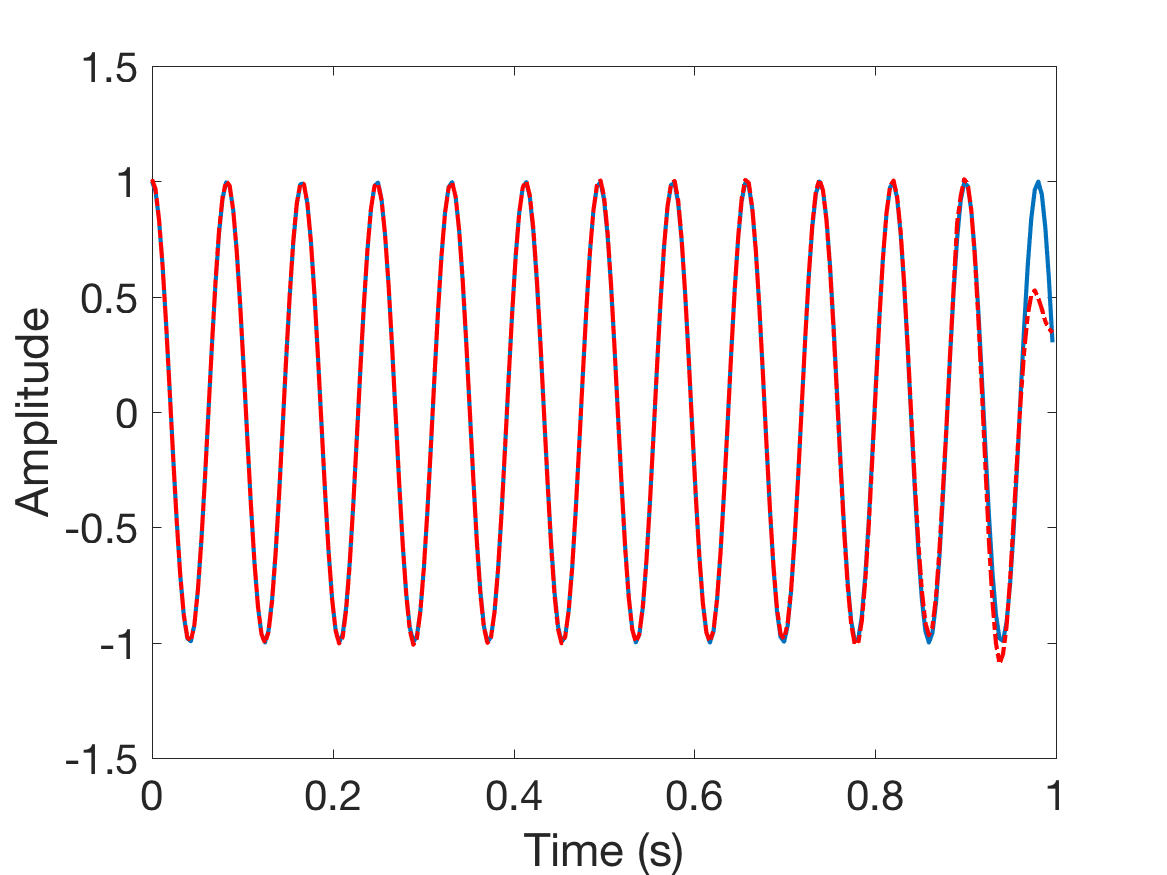}}
\quad 
&\resizebox{3.0in}{2.0in} {\includegraphics{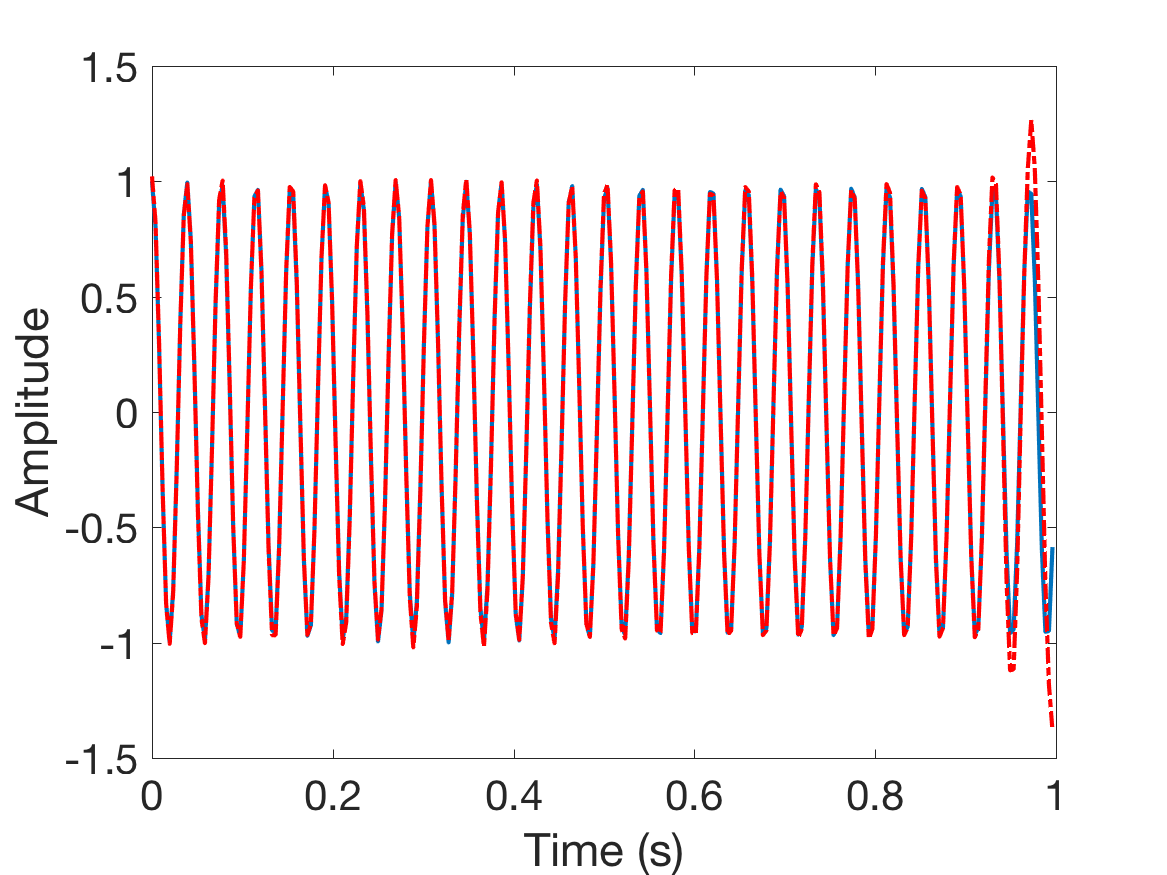}}
	\\
	\resizebox{3.0in}{2.0in}{\includegraphics{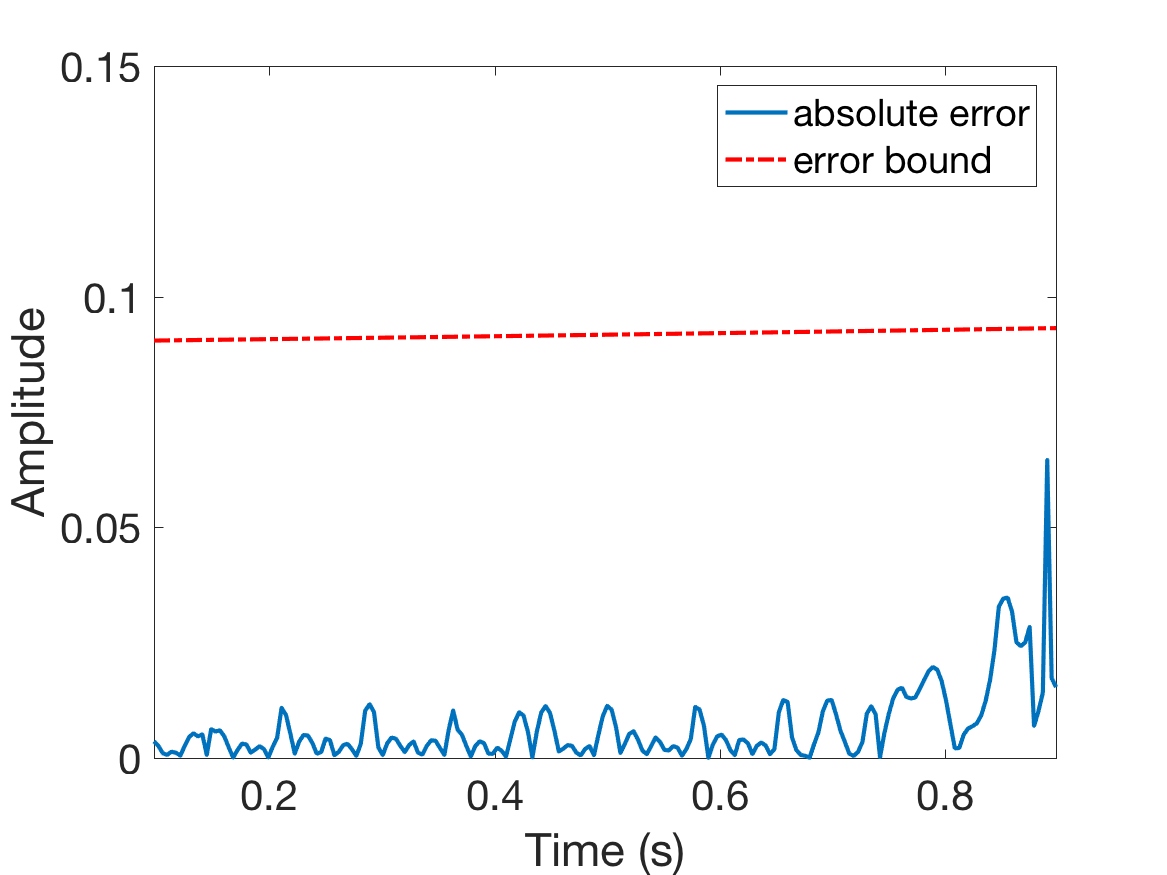}}
\quad 
&\resizebox{3.0in}{2.0in} {\includegraphics{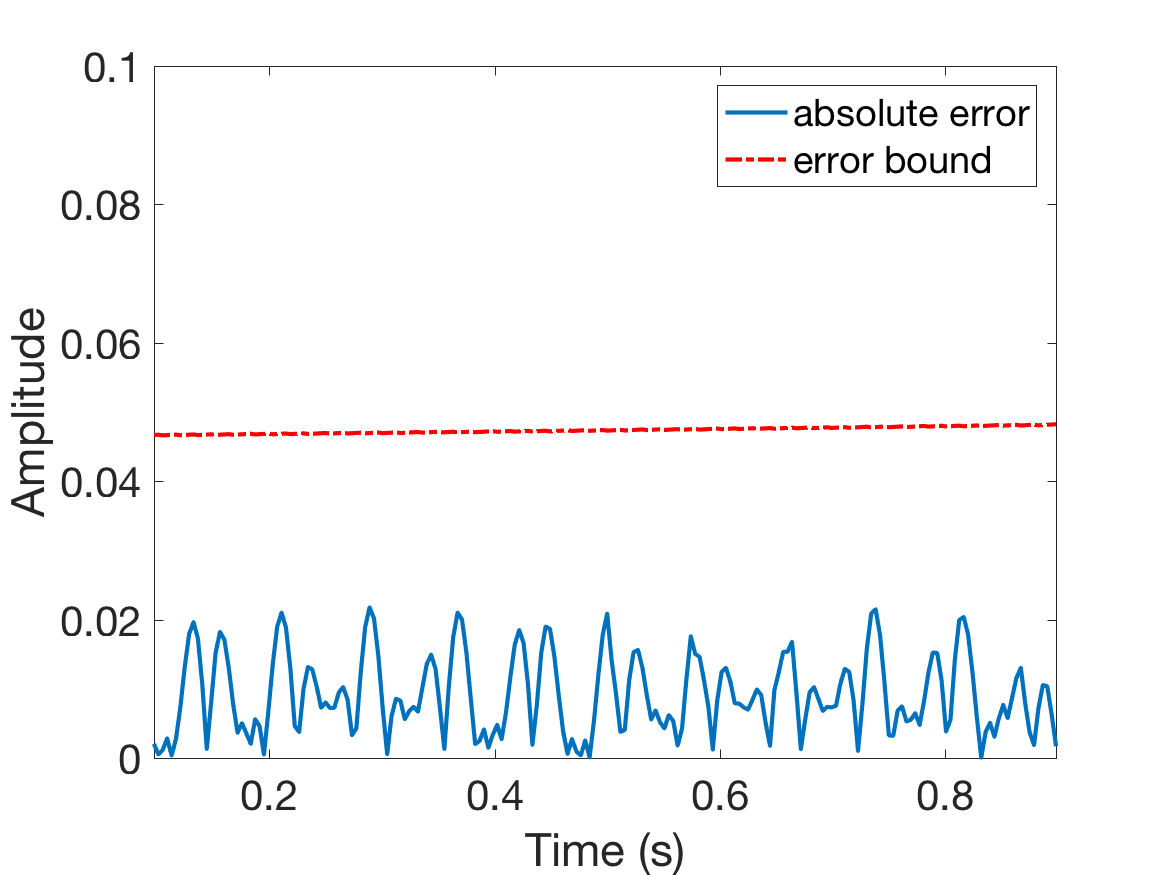}}
	\end{tabular}
	\caption{\small Example of two-component signal $x(t)$ in \eqref{two_chirps1}. 
		Top:  Waveform;  
		Middle-left:  $x_1(t)$ and recovered $x_1(t)$ (red dot-dash line); Middle-right:  $x_2(t)$ and recovered $x_2(t)$ (red dot-dash line);  
		Bottom-left: Absolute recovery error for $x_1$ and error bound $\wt{\rm bd}_1$; 
	Bottom-right: Absolute recovery error for $x_2$ and error bound $\wt{\rm bd}_2$.}
	\label{fig:two_chirp_signal1}
\end{figure}

\begin{example}
\label{example1}
Let $x(t)$ be a two-component linear frequency modulation signal given by  
\begin{equation}
\label{two_chirps1}
x(t):=x_1(t)+x_2(t)
= \cos \left(2\pi(12t+0.5t^2/2)\right)+ \cos\left(2\pi(26t-0.5t^2/2)\right), \quad t\in [0, 1]. 
\end{equation}
\clearpage 
\n The number of sampling points is $N=256$ and the sampling rate is 256Hz.
The instantaneous frequencies of  $x_1(t)$ and $x_2(t)$ are  $\phi'_1(t)=12+0.5t$ and $\phi'_2(t)=26-0.5t$, respectively.
Hence, $x(t)\in \cC_{\vep_1, \vep_2}$ with $\vep_1=0, \vep_2=0.5$. 
In Fig.\ref{fig:two_chirp_signal1}, we show  the waveform of $x(t)$. 

We let $\mu=1$, and choose $\gs(b)$ to be 
$$
\gs_1(b):=\frac \ga\mu \frac{\phi'_2(b)+\phi'_1(b)}{\phi'_2(b)-\phi'_1(b)}.
$$ 
We set $\tau_0=1/20, \wt \ep_1=0.01$ and $\wt \ep_3=\frac 12(\phi'_2(b)-\phi'_1(b)$. 
We show the recovered  $x_1(t), x_2(t)$ in the middle row of Fig.\ref{fig:two_chirp_signal1}. 
The absolute recovery errors {\rm (}the quantity on the left-hand side of   \eqref{reconstr_1st}{\rm )} for $x_1$ and $x_2$ and the error bounds  $\wt{\rm bd}_1$ and $\wt{\rm bd}_2$ 
are provided in the bottom row of Fig.\ref{fig:two_chirp_signal1}. Observe from the middle row of Fig.\ref{fig:two_chirp_signal1} that the recovery errors are small except near boundary points $t=0$ and $t=1$ due to the boundary issue. Hence,  we show the errors for $t\in [0.1, 0.9]$ in the bottom row of Fig.\ref{fig:two_chirp_signal1}. 
\end{example}

\section{Analysis of 2nd-order adaptive WSST}

In this section we consider multicomponent signals $x(t)$ of \eqref{AHM}
satisfying 
the following conditions:
\begin{eqnarray}
 \label{cond_basic_2nd} && A_k(t)\in C^2(\R)\cap L_\infty(\R), \phi_k(t)\in C^3(\R),
\phi^{\gp\gp}_k(t) \in L_\infty(\R),
\end{eqnarray}
We also assume each $x(t)$ is well approximated locally by linear chirp signals of \eqref{def_chip_At}
with $A_k'(t)$ and $\phi_k^{(3)}(t)$  small:
\begin{equation}
\label{condition2}
|A_k'(t)|\le \vep_1, \; |\phi_k^{(3)}(t)|\le \vep_3, \; t\in \R, \; 1\le k\le K,
\end{equation}
for some small positive numbers $\vep_1, \vep_3$. More precisely, write $x(b+at)$ as
\begin{equation}
\label{xm_xr}
x(b+a t)=x_{\rm m}(a,b,t)+x_{\rm r}(a,b,t),
\end{equation}
where
\begin{equation}
\label{xm}
x_{\rm m}(a,b,t):=\sum_{k=1}^K x_k(b)e^{i2\pi (\phi_k'(b) at+\frac 12\phi''_k(b) (at)^2) }
\end{equation}

\begin{eqnarray}
\label{xr}
&&x_{\rm r}(a,b,t):=\sum_{k=1}^K \Big\{ (A_k(b+a t)-A_k(b))e^{i2\pi \phi_k(b+a t)}\\
\nonumber &&\qquad \qquad
+x_k(b)e^{i2\pi (\phi_k'(b) at+\frac 12\phi''_k(b)(at)^2) } \big(
e^{i2\pi (\phi_k(b+a t)-\phi_k(b)-\phi_k'(b) at- \frac 12\phi''_k(b) (at)^2)}-1\big)\Big\}.
\end{eqnarray}
By condition \eqref{condition2}, we have  $|A_k(b+a t)-A_k(b)|\le \vep_1 a |t|$
and
$$
|e^{i2\pi (\phi_k(b+a t)-\phi_k(b)-\phi_k'(b) at- \frac 12\phi''_k(b) (at)^2)}-1|\le
2\pi \frac 16 \sup_{\eta \in \R}|\phi^{(3)}_k(\eta) (a t)^3|
\le \frac \pi 3 \vep_3 a^3 |t|^3.
$$
Thus,
\begin{equation}
\label{xr_est}
|x_{\rm r}(a,b,t)|\le \vep_1K a|t|+\frac \pi 3 \vep_3 a^3 |t|^3 \sum_{k=1}^K A_k(b).
\end{equation}
Therefore, $x_{\rm m}(a, b, t)$ approximates $x(b+at)$ well if $\vep_1, \vep_3$ are small. Note that
$x_{\rm m}(a, b, t)$ is a linear combination of linear chirps with variable $t$.

Next we consider the approximation of $\wt W_x(a, b)$ when $x(b+at)$ is approximated by $x_{\rm m}(a,b,t)$.
With \eqref{xm_xr}, we have
\begin{eqnarray}
\nonumber
\wt W_x(a, b)\hskip -0.6cm &&=\sum_{k=1}^K \int_\R x_k(b+a t)\frac 1{\gs(b)}g\Big(\frac t{\gs(b)}\Big) e^{-i2\pi \mu t}dt\\
\label{CWT_approx0}
&&=\sum_{k=1}^K \int_\R x_k(b)e^{i2\pi (\phi_k'(b) a t +\frac12\phi''_k(b) a^2t^2)}\frac 1{\gs(b)}g\Big(\frac t{\gs(b)}\Big) e^{-i2\pi \mu t}dt +\err_0,
\end{eqnarray}
where
\begin{eqnarray}
\label{def_err0}
&&\err_0:= \int_\R x_{\rm r}(a,b,t)\frac 1{\gs(b)}g\Big(\frac t{\gs(b)}\Big) e^{-i2\pi \mu t}dt.
\end{eqnarray}
For given $a, b$, we use $G_k(\xi)$ to denote the Fourier transform of $e^{i\pi \phi''_k(b)a^2\gs^2(b) t^2}g(t)$, namely,
$$
G_k(\xi):={\cal F}\Big(e^{i\pi \phi''_k(b) a^2 \gs^2(b)t^2}g(t)\Big)\big(\xi)=
\int_{\R} e^{i\pi\gs^2(b) \phi''_k(b) a^2 t^2}g(t) e^{-i2\pi \xi t}dt,
$$
where ${\cal F}$ denotes the Fourier transform. Note that $G_k(\xi)$ depends on $a, b$ also if $\phi''_k(b)\not=0$. We drop $a, b$ in $G_k$
for simplicity. Thus we have
\begin{eqnarray}
\label{CWT_approx}
\wt W_x(a, b)=\sum_{k=1}^K x_k(b)  
G_k\big(\gs(b)(\mu-a \phi_k'(b) )\big)+\err_0.
\end{eqnarray}

Note that to distinguish the different types of the remainders for the expansion of  $\wt W_x(a, b)$ resulted from different local approximations for $x_k(b+a t)$,
in this section we use \lq\lq{}$\err$\rq\rq{}, which means residual,  to denote  the remainder  for the expansion of  $\wt W_x(a, b)$  in \eqref{CWT_approx0}. By \eqref{xr_est}, we have the following estimate for $\err_0$:
\begin{eqnarray*}
|\err_0|\hskip -0.6cm&&\le   \int_\R  K \vep_1 a | t| \frac 1{\gs(b)}\Big|g\Big(\frac t{\gs(b)}\Big)\Big| dt+
\int_\R \frac \pi3 \vep_3 a^3  | t|^3 \sum_{k=1}^K A_k(b)  \frac 1{\gs(b)}\Big|g\Big(\frac t{\gs(b)}\Big)\Big|
 dt \\
 &&=K \vep_1 I_1 a \gs(b)  +\frac \pi 3 \vep_3 I_3 a^3 \gs^3(b) \sum_{k=1}^K A_k(b),
\end{eqnarray*}
where $I_n$ is defined in \eqref{def_In}.
Hence we have
\begin{equation}
\label{err0_est}
|\err_0|\le a \gs(b) \Pi_0(a, b),
\end{equation}
where
$$
\Pi_0(a, b):=K \vep_1  I_1 +\frac \pi 3 \vep_3 I_3 a^2 \gs^2(b) \sum_{k=1}^K A_k(b).
$$

\bigskip
By \eqref{err0_est}, we know $|\err_0|$ is small  if $\vep_1, \vep_3$ are small enough.  Hence, in this case  $G_k\big(\gs(b)(\mu-a \phi_k'(b) )\big)$ determines the scale-time zone for $\wt W_{x_k}(a, b)$. 
More precisely, let $0<\tau_0<1$ be a given small number as the threshold. 
Denote
\begin{equation*}
O'_k:=\{(a, b): |G_k\big(\gs(b)(\mu-a\phi_k'(b))\big)| >\tau_0, b\in \R\}.
\end{equation*}
If $|G_k(\xi)|$ is even and decreasing for $\xi\ge 0$. 
Then $O'_k$ can be written as
\begin{equation}
\label{def_Ok2_old}
O'_k=\Big\{(a, b): |\mu -a \phi_k'(b)|< \frac {\ga_k}{\gs(b)}, b\in \R\Big\}.
\end{equation}
where $\ga_k$ is obtained by solving $|G_k(\xi)|
=\tau_0.$ In general $\ga_k=\ga_k(a,b)$ depends on both $b$ and $a$, and it is hard to obtain the explicit expressions for the boundaries of $Q'_k$.
As suggested in \cite{LCJ18}, in this paper, we assume $\ga_k(a,b)$ can be replaced by $\gb_k(a,b)$ with
$\ga_k(a,b)\le \gb_k(a,b)$ such that $O'_k$ defined by \eqref{def_Ok2_old} with  $\ga_k=\gb_k(a,b)$ can be written as
\begin{equation}
\label{def_Ok2}
O_k:=\{(a, b): l_k(b)<a<u_k(b), b\in \R\},
\end{equation}
for some $0<l_k(b)<u_k(b)$, and 
\begin{equation}
\label{def_Ok2_ineq}
\big|G_k\big(\gs(b)(\mu-a\phi_k'(b))\big)\big|\le \tau_0, \; \hbox{for $(a, b)\not\in Q_k$}. 
\end{equation} 
In addition, we will assume the multicomponent signal $x(t)$ is well-separated, that is there is $\gs(b)$ such that 
\begin{equation}
\label{boundary_ineq}
u_k(b) \le l_{k-1}(b), \; b\in \RR, k=2, \cdots, K, 
\end{equation}
or equivalently
\begin{equation}
\label{cond_no_overlapping}
O_k\cap O_{\ell}=\emptyset, \quad k\not=\ell.
\end{equation}

Next we consider the case that $g$ is the Gaussian function defined by  \eqref{def_g} as an example to illustrate our approach.  One can obtain for this $g$ (see \cite{LCJ18}),
\begin{equation}
\label{CWT_LinearChip}
G_{k}(u)=\frac {1}{\sqrt{1-i2\pi \phi''_k(b)a^2\gs^2(b)}}\;
e^{-\frac{2\pi^2 u^2}{1+(2\pi \phi''_k(b)a^2\gs^2(b))^2} (1+i2\pi \phi''_k(b)a^2\gs^2(b))}.
\end{equation}
Thus
\begin{equation}
\label{abs_Gk}
|G_k(u)|=\frac 1{\big(1+(2\pi \phi''_k(b)a^2\gs^2(b))^2\big)^{\frac 14}}\;
e^{-\frac{2\pi^2}{1+(2\pi \phi''_k(b)a^2 \gs^2(b))^2}u^2}.
\end{equation}
Therefore, in this case,
assuming $\tau_0 (1+  (2\pi \phi''_k(b)a^2 \gs^2(b))^2)^{\frac 14}\le 1$ (otherwise, $|G_k(u)|< \tau_0$ for any $u$),
\begin{equation*}
\ga_k=\ga \sqrt{1+(2\pi \phi''_k(b) a^2 \gs^2(b))^2}
\; \frac 1{2\pi}\sqrt{2\ln (\frac 1{\tau_0})-\frac 12 \ln(1+(2\pi \phi''_k(b)a^2 \gs^2(b))^2)}.
\end{equation*}
Authors of \cite{LCJ18} replaced $\ga_k$ by
$$
\gb_k=\ga\big(1+ 2\pi |\phi''_k(b)|a^2 \gs^2(b)\big),
$$
where $\ga= \frac 1{2\pi}\sqrt{2\ln (1/{\tau_0})}$ as defined by \eqref{def_ga}. 
Since $\ga_k\le \gb_k$, we know \eqref{def_Ok2_ineq} holds.  
That is $\wt W_{x_k}(a, b)$ lies within the scale-time zone: 
   \begin{equation*}
 \Big\{(a, b): |\mu -a \phi_k'(b)|< \frac {\ga}{\gs(b)}
  \Big(1+ 2\pi |\phi''_k(b)| a^2 \gs^2(b)\Big), b\in \R\Big\},
\end{equation*}
which can written as \eqref{def_Ok2} with (see \cite{LCJ18})
\begin{equation}
\label{def_ulk}
\begin{array}{l}
u_k(b) =\frac {2(\mu+\frac \ga{\gs(b)})}{\phi'_k(b)+\sqrt{\phi'_k(b)^2-8\pi \ga (\ga+\mu \gs(b))|\phi''_k(b)|}},\\
 \\
l_k(b) =\frac {2(\mu-\frac \ga{\gs(b)})}{\phi'_k(b)+\sqrt{\phi'_k(b)^2+8\pi \ga (\mu \gs(b)-\ga)|\phi''_k(b)|}}.
\end{array}
\end{equation}
It was shown in \cite{LCJ18} that if 
\begin{eqnarray}
\label{sep_cond1}
&& 4\ga \sqrt{\pi} \sqrt {|\phi''_k(b)|+|\phi''_{k-1}(b)|}\le \phi'_k(b)-\phi'_{k-1}(b), \quad k=2, \cdots, K, 
 \end{eqnarray}
then  \eqref{boundary_ineq} holds if and only if $\gs$ satisfies
\begin{equation}
\label{gs_ineq}
\frac{\gb_k(b)-\sqrt{\Upsilon_k(b)}}{2\ga_k(b)}\le \gs\le \frac{\gb_k(b)+\sqrt{\Upsilon_k(b)}}{2\ga_k(b)},
\end{equation}
where 
\begin{eqnarray*}
\label{gak}
&&\ga_k(b):=2\pi \ga \mu (|\phi''_k(b)|+|\phi''_{k-1}(b)|)^2,\\
\label{gbk}
&&\gb_k(b):=\big(\phi'_k(b)|\phi''_{k-1}(b)|+\phi'_{k-1}(b)|\phi''_{k}(b)|\big) \big(\phi'_k(b)-\phi'_{k-1}(b)\big)+4\pi \ga^2  \big(\phi''_k(b)^2-\phi''_{k-1}(b)^2\big),\\
\label{ggak}
&&\gga_k(b):=\frac{\ga}\mu\Big\{
\big(\phi'_k(b)|\phi''_{k-1}(b)|+\phi'_{k-1}(b)|\phi''_{k}(b)|\big)\big(\phi'_k(b)+\phi'_{k-1}(b)\big)+2\pi \ga^2  \big(|\phi''_k(b)|-|\phi''_{k-1}(b)|\big)^2\Big\}, 
\end{eqnarray*}
and 
\begin{eqnarray*}
&&\Upsilon_k(b):=\gb_k(b)^2-4\ga_k(b)\gga_k(b)\\
&&=\big(\phi'_k(b)|\phi''_{k-1}(b)|+\phi'_{k-1}(b)|\phi''_{k}(b)|\big)^2\Big\{\big(\phi'_k(b)-\phi'_{k-1}(b)\big)^2-16\pi \ga^2 \big(|\phi''_k(b)|+|\phi''_{k-1}(b)|\big)
\Big\}. 
\end{eqnarray*}
Thus  \cite{LCJ18} calls \eqref{sep_cond1} and \eqref{sep_cond2} below the well-separated conditions: 
\begin{eqnarray}
 \label{sep_cond2}
 && \max\Big\{\frac \ga\mu, \frac{\gb_k(b)-\sqrt{\Upsilon_k(b)}}{2\ga_k(b)}: 2\le k\le K\Big\}\le \min_{2\le k\le K}\Big\{\frac{\gb_k(b)+\sqrt{\Upsilon_k(b)}}{2\ga_k(b)}\Big\}. 
 \end{eqnarray}
\cite{LCJ18} suggests to choose $\gs(b)$ to be $\gs_2(b)$ defined by 
\begin{equation}
\label{def_gs2}
\gs_2(b):=\left\{
\begin{array}{ll}
\max\Big\{\frac\ga\mu, \frac{\gb_k(b)-\sqrt{\Upsilon_k(b)}}{2\ga_k(b)}: \; 2\le k\le K\Big\}, &\hbox{if $|\phi''_k(b)|+|\phi''_{k-1}(b)|\not=0$}, \\
& \\
 \max\Big\{\frac \ga\mu \frac{\phi'_k(b)+\phi'_{k-1}(b)}{\phi'_k(b)-\phi'_{k-1}(b)}: \; 2\le k\le K\Big\}, &\hbox{if $\phi''_k(b)=\phi''_{k-1}(b)=0$.}
\end{array}
\right.
\end{equation}

\bigskip
In the following we assume  $x(t)$ given by \eqref{AHM} satisfy \eqref{freq_resolution_adp} and  \eqref{cond_basic_2nd},  and that the adaptive CWTs 
$\wt W_{x_k}(a,b)$ of its components  with a window function $g \in \cS$ 
lie within scale-time zones $Q_k$ in the sense that 
\eqref{def_Ok2_ineq} holds and each  $Q_k$ is  given by \eqref{def_Ok2}. 
In addition, we assume $x(t)$ is well-separated, that is there is $\gs(b)$ such that \eqref{cond_no_overlapping} holds. Let $\cE_{\vep_1, \vep_3}$ denote the set of such multicomponent signals 
$x(t)$ satisfying \eqref{condition2}. 

\bigskip

Next we introduce more notations to describe our main theorems on the 2nd-order adaptive WSST. For $j\ge 0$, denote
\begin{eqnarray}
\label{def_Gjk}
G_{j,k}(a, b)\hskip -0.6cm &&:=\int_{\R} e^{i2\pi (\phi_k'(b) a t +\frac12\phi''_k(b) a^2 t^2)}\frac {t^j}{\gs(b)^{j+1}}g\Big(\frac t{\gs(b)}\Big) e^{-i2\pi \mu t}dt\\
\nonumber &&={\cal F}\Big(e^{i\pi \phi''_k(b) a^2\gs^2(b) t^2}t^j g(t)\Big)\big(\gs(b)(\mu-a \phi_k'(b))\big)\big).
\end{eqnarray}
Clearly 
$$
G_{0, k}(a, b)=G_k\big(\gs(b)(\mu-a \phi_k'(b) )\big).
$$
We also denote
\begin{eqnarray*}
\label{def_Bk}
&&B_k(a, b):=\sum_{\ell\not=k} x_\ell(b) \big(\phi'_\ell(b)-\phi_k'(b)\big)G_{0, \ell}(a, b),
\\ 
\label{def_Dk} &&
D_k(a, b):=\sum_{\ell\not=k} x_\ell(b) \big(\phi''_\ell(b)-\phi''_k(b)\big)G_{1, \ell}(a, b),\\
\label{def_Ek}
&&E_k(a, b):=\sum_{\ell\not=k} x_\ell(b) \big(\phi'_\ell(b)-\phi_k'(b)\big)\big(\phi'_\ell(b)G_{1, \ell}(a,b)+
\phi''_\ell(b) a\gs(b) G_{2, \ell}(a,b)\big),\\
&& F_k(a, b):=\sum_{\ell\not=k} x_\ell(b) \big(\phi''_\ell(b)-\phi''_k(b)\big)\big(\phi'_\ell(b)G_{2, \ell}(a,b)+
\phi''_\ell(b) a\gs(b) G_{3, \ell}(a,b)\big),
\end{eqnarray*}
and denote 
\begin{equation}
\label{def_Mlk}
M_{\ell, k}(b):=\int_{\{a: \; (a, b)\in O_k\}}\left| G_{0, \ell}(a, b)\right| \frac{da}a
= \int_{l_k(b)}^{u_k(b)} \big| G_\ell \big(\gs(b)(\mu-a \phi'_\ell(b))\big) \big| \frac{da}a.
\end{equation}

\bigskip

Recall that $\wt W^{g_j}_x(a, b), j=1, 2, 3$ and $\wt W^{g'}_x(a, b)$ denote respectively the adaptive CWTs defined by \eqref{def_CWT_para} with $g$ replaced by $g_j$ and $g'$, where $g_j$ are defined by
\eqref{def_psi_j}. Expand $\wt W^{g_j}_x(a, b), j=1, 2, 3$ and $\wt W^{g'}_x(a, b)$ as $\wt W_x(a, b)$ in \eqref{CWT_approx0}, and let $\err_1$, $\err_2$, $\err_1'$ and $\err_0'$ be the corresponding residuals.
Then $\err_1$, $\err_2$, $\err_1'$, and $\err_0'$ 
are given as $\err_0$ in \eqref{def_err0} with $g(t)$ replaced respectively by $tg(t)$,  $t^2g(t)$,  $tg'(t)$, and $g'(t)$. Thus we have the estimates for these residuals similar to \eqref{err0_est}. More precisely, we have
\begin{equation}
\label{err_est_various}
|\err_1|\le a\gs(b) \Pi_1(a, b),  |\err_2|\le a\gs(b) \Pi_2(a, b), |\err_0'|\le a\gs(b) \wt \Pi_0(a, b), 
|\err_1'|\le a\gs(b) \wt \Pi_1(a, b),
\end{equation}
where
\begin{eqnarray*}
&&\Pi_1(a, b):=K \vep_1  I_2 +\frac \pi 3 \vep_3 I_4 a^2 \gs^2(b) \sum_{k=1}^K A_k(b), \\
&& \Pi_2(a, b):=K \vep_1  I_3 +\frac \pi 3 \vep_3 I_5 a^2 \gs^2(b) \sum_{k=1}^K A_k(b), \\
&&\wt \Pi_0(a, b):=K \vep_1  \wt I_1 +\frac \pi 3 \vep_3 \wt I_3 a^2 \gs^2(b) \sum_{k=1}^K A_k(b), \\
&&\wt \Pi_1(a, b):=K \vep_1  \wt I_2 +\frac \pi 3 \vep_3 \wt I_4 a^2 \gs^2(b) \sum_{k=1}^K A_k(b)
\end{eqnarray*}
with $I_n$ and $\wt I_n$ defined by \eqref{def_In} and \eqref{def_tIn} respectively.

\bigskip
Next we provide Theorem \ref{theo:main} on the 2nd-order adaptive WSST. 
The proof of Part (b) of  Theorem \ref{theo:main} is based on the following three lemmas whose proofs are postponed to Appendix C. The residuals $\Err_1, \Err_2$ in these lemmas are defined as 
\begin{eqnarray}
\label{def_Err1}
\label{def_Err2}
\Err_1:=\Err_{1,1}+\Err_{1,2}, \; \Err_2:=\Err_{2,1}+\Err_{2,2},
 \end{eqnarray}
 where 
 \begin{eqnarray*}
&& \Err_{1,1}:=i2\pi B_k(a, b)+i 2\pi a\gs(b)D_k(a, b), \\
&& \Err_{1,2}:={i2\pi} \Big(\frac \mu a -\phi_k'(b)\Big)\err_0 -  \frac{ \err_0'}{a\gs(b)}-i2\pi \phi''_k(b)a \gs(b)\err_1,  \\
&& \Err_{2, 1}:=-4\pi^2 \gs(b) E_k(a, b)+i2\pi\gs(b) D_k(a,b)-4\pi^2 a\gs^2(b)F_k(a, b)
\\
&&\Err_{2, 2}:= \frac{ i 2\pi}{a^2}(a\phi^\gp_k(b)-2\mu)  \; (\err_0+\err^\gp_1)
+i2\pi\gs(b) \big(\phi^{\gp\gp}_k(b)-\frac{i2\pi \mu}{a^2}(a \phi_k'(b)-\mu)\big)\; \err_1 \\
&&\qquad \qquad
 -i2\pi\gs(b) \phi^{\gp\gp}_k(b) \big(i2\pi \mu \gs(b) \; \err_2 -\err^\gp_2 \big)+
\frac 1{a^2\gs(b)}(2\; \err'_0 +\err''_1) 
 \end{eqnarray*}
with $\err^\gp_2 $ and $\err''_1$ are the errors defined by  \eqref{def_err0} with $g(t)$ replaced by $t^2 g^\gp(t)$ and  $t g^{\gp\gp}(t)$ respectively.

\begin{lem}
\label{lem:lem1} Let $\Err_1$ be the quantity defined by \eqref{def_Err1}. Then
\begin{equation}
\label{result_lem1}
\pd_b \wt W_x(a, b)=\Big(i2\pi \phi_k'(b)-\frac {\gs'(b)}{\gs(b)}\Big)\wt W_x(a, b)+
i2\pi \phi_k''(b) a \gs(b)\wt W^{g_1}_x(a, b)- \frac {\gs'(b)}{\gs(b)}\wt W^{g_3}_x(a, b)+\Err_1.
\end{equation}
\end{lem}

\begin{lem}
\label{lem:lem2} Let $\Err_2$ be the quantity defined by \eqref{def_Err2}. Then $\pd_a \Err_1=\Err_2$, and 
\begin{eqnarray}
\label{result_lem2}
&& \pd_a \pd_b \wt W_x(a, b)=\Big(i2\pi \phi_k'(b)-\frac {\gs'(b)}{\gs(b)}\Big)\pd_a \wt W_x(a, b)\\
\nonumber
&&\qquad \qquad
+ i2\pi \phi_k''(b)\gs(b) \big( \wt W^{g_1}_x(a, b)+a \pd_a \wt W^{g_1}_x(a, b) \big) - \frac {\gs'(b)}{\gs(b)}\pd_a \wt W^{g_3}_x(a, b)+\Err_2.
\end{eqnarray}
\end{lem}

\begin{lem}
\label{lem:lem3} Let $R_0(a,b)$ be the quantity defined by \eqref{def_R0}. Then 
for $(a, b)$ satisfying $\wt W_x(a, b)\not =0$ and
$\frac{\partial}{\partial a}\Big( \frac {a \wt W^{g_1}_x(a, b)}{\wt W_x(a, b)}\Big)\not=0$, we have
\begin{equation}
\label{result_lem3}
R_0(a, b)=i2\pi \gs(b)\phi^{\prime\prime}_k(b)+\Err_3,
\end{equation}
where 
\begin{equation}
\label{Err3}
\Err_3:=\frac {\wt W_x(a, b) \; \Err_2-\pd _a \wt W_x(a, b) \; \Err_1 }{\wt W_x(a, b) \wt W^{g_1}_x(a, b)+a \wt W_x(a, b)\pd _a
\wt W^{g_1}_x(a, b) -a \wt W^{g_1}_x(a, b)\pd _a
\wt W_x(a, b)}
\end{equation}
with $\Err_1$ and $\Err_2$ defined by \eqref{def_Err1}. 
\end{lem}

\bigskip

\begin{theo}
\label{theo:main} Suppose $x(t)\in \cE_{\vep_1, \vep_3}$ for some small $\vep_1, \vep_3>0$. Then we have the following.

{\rm (a)} Suppose $\wt \vep_1$ satisfies  $\wt \vep_1 \ge
a_2(b)\gs(b) \Pi_0(a_2(b), b)+\tau _0 \sum_{k=1}^K A_k(b)$. Then for $(a, b)$ with $|\wt W_x(a, b)|>\wt \vep_1$, there exists $k\in \{1, 2, \cdots, K\}$ such that $(a, b)\in O_k$.

${\rm (b)}$ Suppose $(a, b)$ satisfies $|\wt W_x(a, b)|>\wt \vep_1$,  $|\pd _a \big(a \wt W^{g_1}_x(a, b)/\wt W_x(a, b)\big)|>\wt \vep_2$,  and $(a, b)\in O_k$.
Then
\begin{equation}
\label{transformation_approx_2nd}
\go_x^{\rm 2adp, c}(a, b)-\phi_k'(b)= \Err_4,
\end{equation}
where
\begin{equation*}
\Err_4:=\frac1{i2\pi \wt W_x(a, b)}\big(\Err_1- a \wt W^{g_1}_x(a, b)\Err_3\big).
\end{equation*}
Furthermore,
\begin{equation}
\label{IF_est_Bd1}
|\go_x^{\rm 2adp}(a, b)-\phi_k'(b)|<  \Bd_k,
\end{equation}
where
\begin{equation}
\label{def_Bd1}
\Bd_k:=\sup_{l_k(b) < a <u_k(b)}\Big\{\frac{|\Err_1|}{2\pi \wt \vep_1}+
\frac1{2\pi \wt \vep_1^3 \wt \vep_2}
a |\wt W^{g_1}_x(a, b)| \big(\wt \vep_1 | \; \Err_2|+
|\pd _a \wt W_x(a, b)|\;  |\Err_1| \big)\Big\}.
\end{equation}

{\rm (c)} Suppose that $\wt \vep_1$ satisfies the condition in part {\rm (a)} and  $\max\limits_{1\le \ell\le K}\{\Bd_\ell\} \le \frac 12 L_k(b)$,
where $L_k(b)$ is defined by \eqref{def_Lk}. 
Then for any $\wt \vep_3=\wt \vep_3(b)>0$ satisfying  $\max\limits_{1\le \ell\le K}\{\Bd_\ell\} \le \wt \vep_3\le\frac 12  L_k(b)$, we have
\begin{equation}
 \label{reconstr}
 \Big| \lim_{\gl\to 0} \frac1{c^k_\psi(b)}\int_{|\xi-\phi_k'(b)|<\wt \vep_3} T_{x, \wt \vep_1, \wt \vep_2}^{\rm 2adp, \gl}(\xi, b)d\xi -x_k(b) \Big|\le \frac 1{|c^k_\psi(b)|}\; \wt \Bd_k,
\end{equation}
where
\begin{equation}
\label{def_ck_psi}
c^k_\psi(b):=\int_{l_k(b)}^{u_k(b)} G_k\big(\gs(b)(\mu-a \phi'_k(b))\big) \frac{da}a
\end{equation}
and $\wt \Bd_k:=\wt \Bd'_k+\wt \Bd''_k$ with
\begin{eqnarray}
\label{def_Bd2}
&& \wt \Bd'_k :=  \wt \ep_1 \ln \frac {u_k(b)}{l_k(b)}
 +\gs(b) K \vep_1  I_1(u_k-l_k) \\
\nonumber &&\qquad \qquad +\frac \pi 9 \vep_3 I_3 (u_k-l_k)^3 \gs^3(b) \sum\limits_{j=1}^K A_j(b)
  +\sum\limits_{\ell\not=k }A_\ell(b)  M_{\ell, k}(b) \\
\nonumber &&  \wt \Bd''_k :=\frac{A_k(b)}{l_k(b)} \|g\|_1 |U_b|+ \gs(b) K \vep_1  I_1(u_k-l_k) \\
&&\nonumber\qquad  \qquad +\frac \pi 9 \vep_3 I_3 (u_k-l_k)^3 \gs^3(b) \sum\limits_{j=1}^K A_j(b)
+\sum\limits_{\ell\not=k }A_\ell(b) M_{\ell, k}(b) 
\end{eqnarray}
and $|U_b|$ denoting the Lebesgue measure of the set $U_b$:
 \begin{equation}
 \label{def_Zt}
 U_b:= \big\{a: \; (a, b)\in O_k, |W_x(a, b)|>\wt \vep_1, \big| \partial _a\big(a{\wt W^{g_1}_x(a, b)}/{\wt W_x(a, b)}\big)\big| \le \wt \vep_2\big\}.
 \end{equation}
\end{theo}

\bigskip
Note that the error bound $\wt \Bd_k$ for the component recovery \eqref{reconstr} also depends on  the Lebesgue measure of the set $U_b$. This makes sense since $T_{x, \wt \vep_1, \wt \vep_2}^{\rm 2adp, \gl}(\xi, b)$ defined by \eqref{def_adp2ndWSST}
takes the integral along the set 
$$
\big\{a>0: \; |\wt W_x(a, b)|>\wt \vep_1, \; |{\partial_a}(a{\wt W^{g_1}_x(a, b)}/{\wt W_x(a, b)})| >\wt \vep_2\big\}, 
$$
namely, $T_{x, \wt \vep_1, \wt \vep_2}^{\rm 2adp, \gl}(\xi, b)$ does not take account of $a$ in  $U_b$. 
Thus only in the case that $|U_b|$ is small, the integral of $T_{x, \wt \vep_1, \wt \vep_2}^{\rm 2adp, \gl}(\xi, b)$ in 
\eqref{reconstr}  can provide accurate component recovery. 

Next we consider another type of 2nd-order WSST $S_{x, \wt \vep_1, \wt \vep_2}^{\rm 2adp,\gl}(\xi, b)$ defined by \eqref{def_adp2ndWSST_3}, where the integral is taken along $\{a>0: |\wt W_x(a, b)|> \wt \vep_1\}$. To this regard, for a given $b\in \R$, denote
\begin{eqnarray}
\label{def_Yt}
&&
V_b:=\big\{a: \; (a, b)\in O_k, |\wt W_x(a, b)|>\wt \vep_1, \big| \partial _a\big(a{\wt W^{g_1}_x(a, b)}/{\wt W_x(a, b)}\big)\big| >\wt \vep_2\big\}.
\end{eqnarray}

\begin{theo}
\label{theo:main3} Suppose $x(t)\in \cE_{\vep_1, \vep_3}$ with a window function $g(t)$ 
for some small $\vep_1, \vep_3>0$. Then besides {\rm (a)} in Theorem \ref{theo:main}, the following hold:

${\rm (b_1)}$ Suppose $(a, b)$ with $a\in V_b$, we have
\begin{equation}
\label{IF_est_Bd1p}
|\go_x^{\rm 2adp}(a, b)-\phi_k'(b)|<  \Bd_1\rq{},
\end{equation}
where
\begin{equation}
\label{def_Bd1p}
\Bd_1\rq{}:=\max_{1\le k\le K}\sup_{a \in V_b}\Big\{\frac{|\Err_1|}{2\pi \wt \vep_1}+ \frac1{2\pi \wt \vep_1^3 \wt \vep_2}
a|\wt W^{g_1}_x(a, b)| \big(|\pd _a \wt W_x(a, b)|\;  |\Err_1| + \wt \vep_1 |\Err_2|\big)\Big\}.
\end{equation}

${\rm (b_2)}$ Suppose $(a, b)$ satisfies $|\wt W_x(a, b)|>\wt \vep_1$  and $(a, b)\in O_k$. Then
\begin{equation}
\label{transformation_approx_1st_chirp}
\go_x^{\rm adp, c}(a, b)-\phi_k'(b)= \phi''_k(b) a\gs(b) \frac{\wt W^{g_1}_x(a, b)}{\wt W_x(a, b)}
+\frac{\Err_1}{i2\pi\wt W_x(a, b)}.
\end{equation}
Thus,  for $a \in U_b$, we have
\begin{equation}
\label{transformation_est_1st_chirp3}
|\go_x^{\rm adp}(a, b)-\phi_k'(b)|<  \Bd_2\rq{}:=\max_{1\le k\le K}\sup_{a \in U_b}\Big\{\frac1{\wt \vep_1}|\phi''_k(b)| a\gs(b)  |\wt W^{g_1}_x(a, b)| +\frac1{2\pi \wt \vep_1} |\Err_1|\Big\}.
\end{equation}
${\rm (c)}$ Suppose that $\wt \vep_1$ satisfies the condition in part {\rm (a)} of Theorem \ref{theo:main}.
In addition, suppose the following two conditions hold: {\rm (i)} $\Bd_1\rq{} \le \frac 12 L_k(b)$, {\rm (ii)} $\Bd_2\rq{} \le \frac 12 L_k(b)$, where $L_k(b)$ is given in \eqref{def_Lk}. 
Then for any $\wt \vep_3=\wt \vep_3(b)>0$ satisfying $\max\{\Bd_1\rq{},\Bd_3\rq{}\}\le \wt \vep_3\le\frac 12  L_k(b)$,
\begin{equation}
 \label{reconstr3}
 \Big| \lim_{\gl\to 0} \frac1{c^k_\psi(b)}\int_{|\xi-\phi_k'(b)|<\wt \vep_3} S_{x, \wt \vep_1, \wt \vep_2}^{\rm 2adp, \gl}(\xi, b)d\xi -x_k(b) \Big|\le \frac1{|c^k_\psi(b)|} \wt \Bd'_k,
\end{equation}
where $c^k_\psi(b)$ is defined by \eqref{def_ck_psi}, and $\wt \Bd'_k$ is defined by \eqref{def_Bd2}.
\end{theo}

The proofs of Theorems \ref{theo:main}  and \ref{theo:main3} will be provided in Appendix B. 

\bigskip 
Compared with \eqref{reconstr},  the integral of $S_{x, \wt \vep_1, \wt \vep_2}^{\rm 2adp, \gl}(\xi, b)$ in 
\eqref{reconstr3}  provides more accurate component recovery. However, in this case there is a restriction on 
$\phi''_k(b)$  on the set $U_b$: $\Bd_2\rq{} \le \frac 12 L_k(b)$.

The error bounds 
$\Bd_k$, $\Bd_1\rq{}$, $\Bd_2\rq{}$ in \eqref{def_Bd1}, 
\eqref{def_Bd1p} and \eqref{transformation_est_1st_chirp3} for instantaneous frequency estimates are determined by $\Err_1$ and $\Err_2$.  From their definitions in \eqref{def_Err1}, 
we know $\Err_1$ and $\Err_2$ are bounded by 
$|B_k(a, b)|$, and/or  $|D_k(a, b)|$, $|E_k(a, b)|$, $|F_k(a, b)|$, and/or $\Pi_j(a,b)$, $\wt \Pi_j(a,b)$ for $j=0, 1, 2$ (refer to \eqref{err_est_various}),  and $\wt{\wt \Pi}_1(a,b)$, where $\wt{\wt \Pi}_1(a,b)$ is defined as $\Pi_1(a,b)$ with $I_2, I_4$ replaced respectively by 
 $$
 \int_\R  t^2 |g^{\gp\gp}(t)| dt, \; \int_\R  t^4 |g^{\gp\gp}(t)| dt. 
 $$
Under decay conditions on $G_k(u)$ and $G_{j, \ell}(a, b)$, $|B_k(a, b)|$, $|D_k(a, b)|$, $|E_k(a, b)|$, $|F_k(a, b)|$ are small for $(a,b)\in O_k$, while $\Pi_j(a,b)$, $\wt \Pi_j(a,b)$,  $\wt{\wt \Pi}_j(a,b)$ 
are small as long as $\vep_1, \vep_3$ are small. Thus $\Err_1$ and $\Err_2$ are small. 
For the component recovery error bounds in \eqref{reconstr} and \eqref{reconstr3}, $M_{\ell, k}(b), \ell\not=k$ are small  if $G_k(u)$ has certain decay. Thus under certain extra conditions, Theorem \ref{theo:main} and \ref{theo:main3} can be stated in the formulation in Corollary \ref{cor:theo_standand_form}.

Finally we consider another example to illustrate the recovery error bounds $\wt \Bd^\gp_k$ in 
\eqref{reconstr3}. 

\begin{example}
\label{example2}
Let $y(t)$ be another  two-component linear frequency modulation signal given by  
\begin{equation}
\label{two_chirps2}
y(t):=y_1(t)+y_2(t)
= \cos \left(2\pi(20t+18t^2/2)\right)+ \cos\left(2\pi(42t+36t^2/2)\right), \quad t\in [0, 1]. 
\end{equation}
Again we set the number of sampling points to be $N=256$ and the sampling rate  256Hz.
The instantaneous frequencies of  $y_1(t)$ and $y_2(t)$ are  $\phi'_1(t)=20+18t$ and $\phi'_2(t)=42+36t$, respectively. Clearly $y_1(t)$ and $y_2(t)$ have fast changing frequencies. 
In Fig.\ref{fig:two_chirp_signal2}, we show  the waveform of $y(t)$. 
\begin{figure}[th]
	\centering
	\begin{tabular}{cc}
		\resizebox{3.0in}{2.0in}{\includegraphics{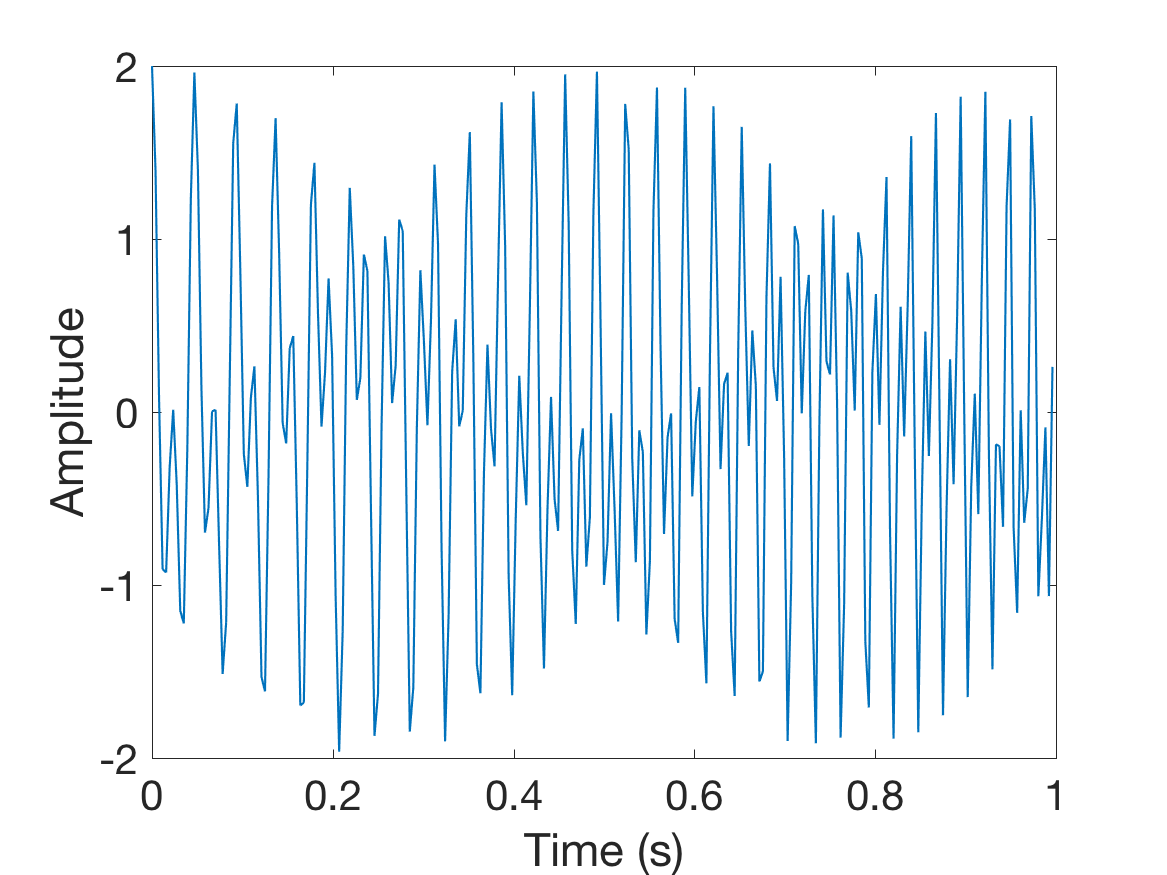}}\quad & \\
\resizebox{3.0in}{2.0in}{\includegraphics{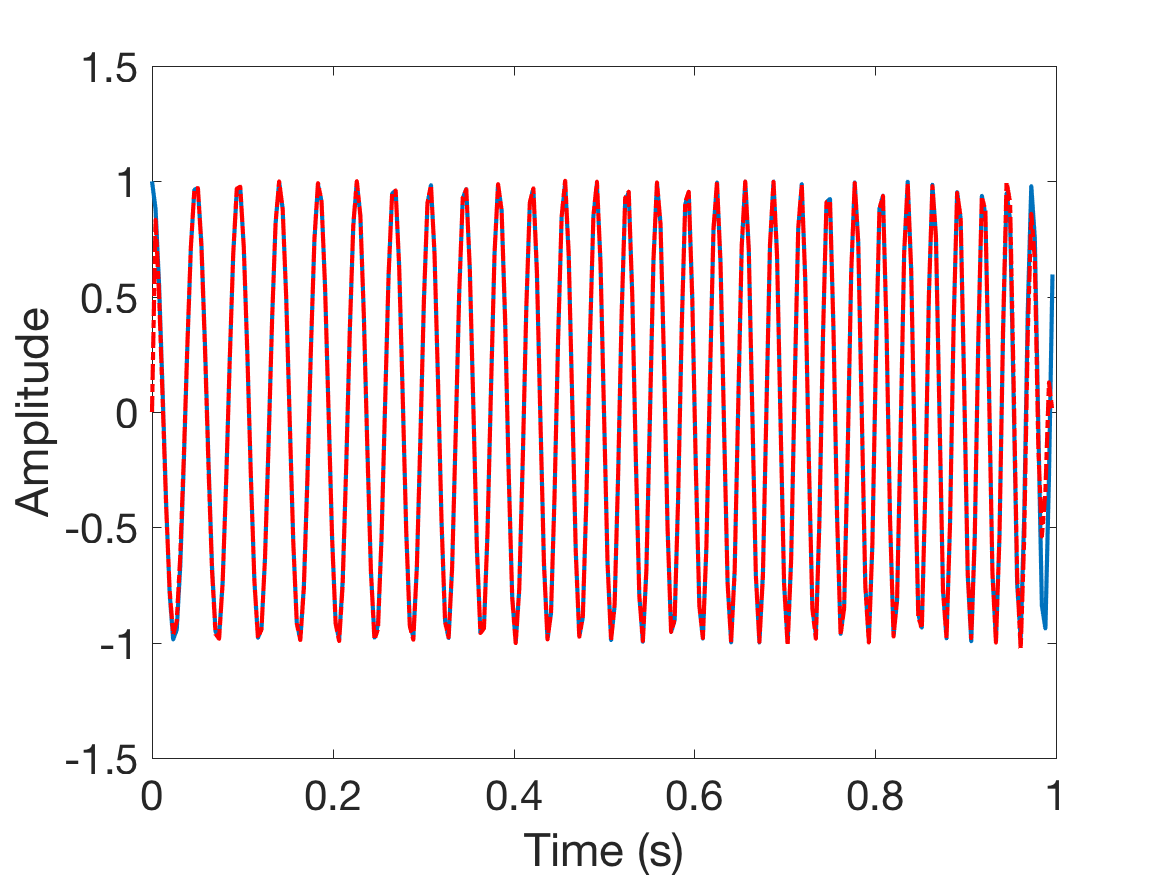}}
\quad 
&\resizebox{3.0in}{2.0in} {\includegraphics{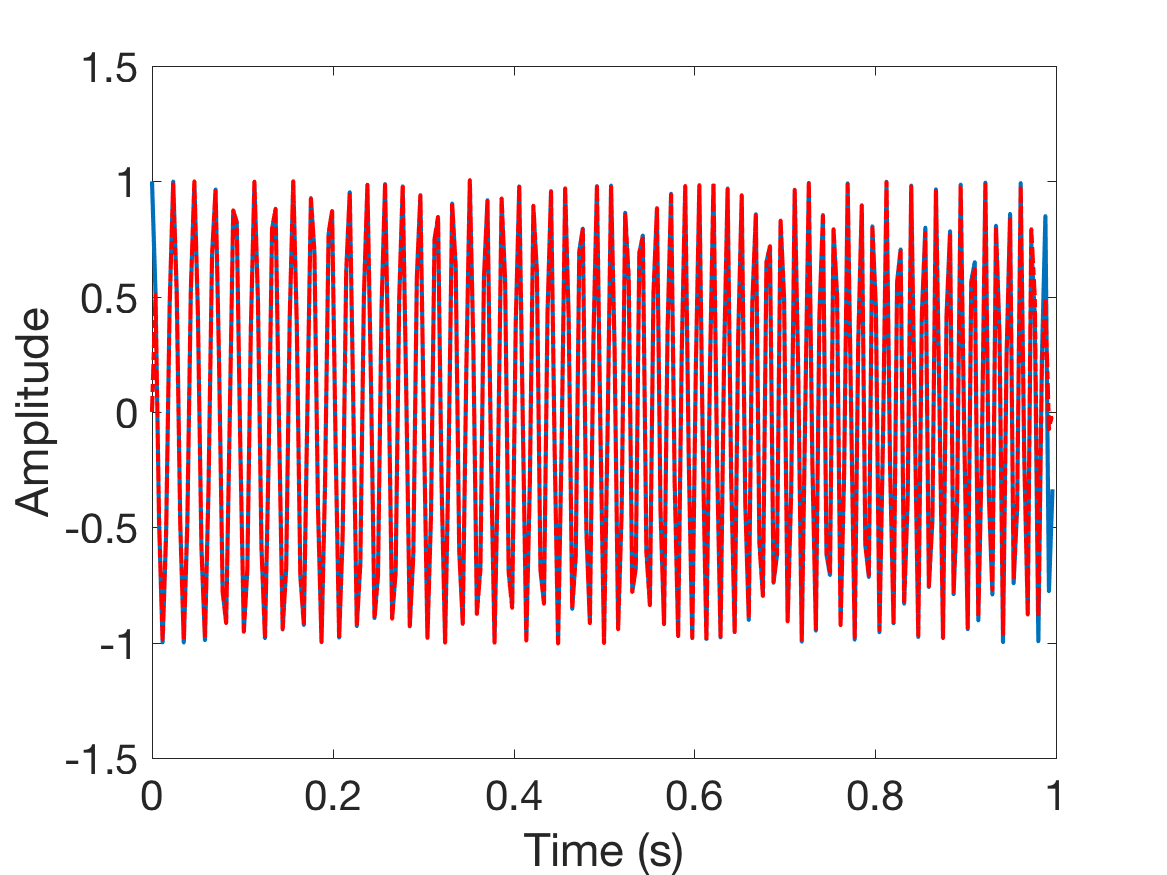}}
	\\
	\resizebox{3.0in}{2.0in}{\includegraphics{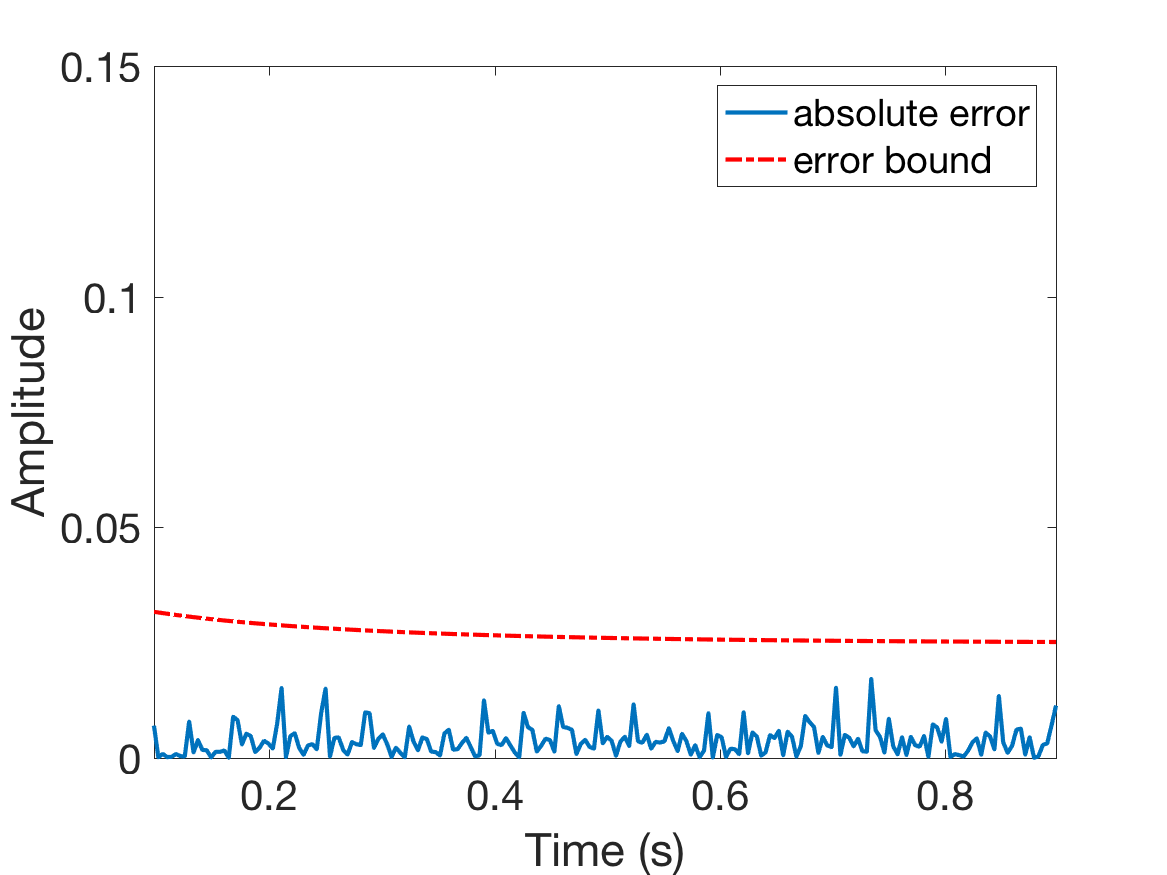}}
\quad 
&\resizebox{3.0in}{2.0in} {\includegraphics{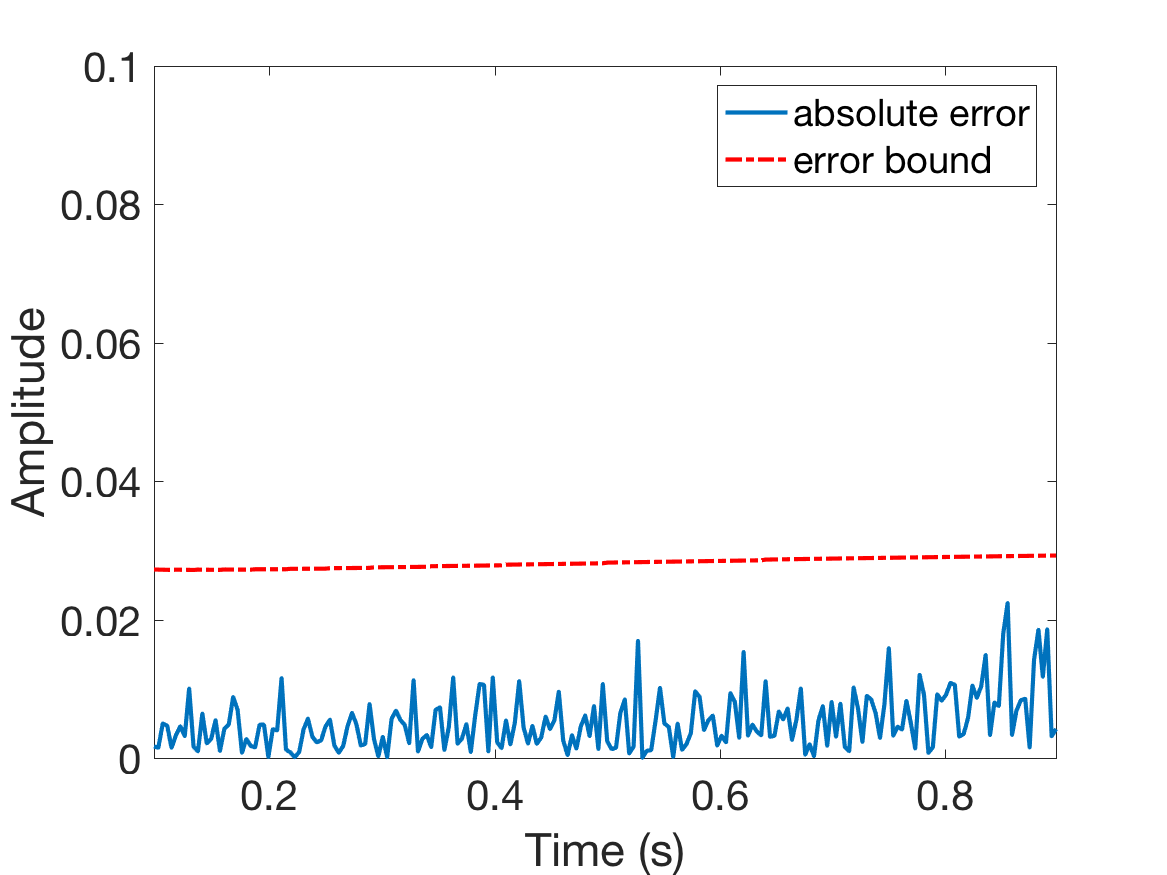}}
	\end{tabular}
	\caption{\small Example of two-component signal $y(t)$ in \eqref{two_chirps2}. 
		Top:  Waveform;  
		Middle-left:  $y_1(t)$ and recovered $y_1(t)$ (red dot-dash line); Middle-right:  $y_2(t)$ and recovered $y_2(t)$ (red dot-dash line);  
		Bottom-left: Absolute recovery error for $y_1$ and error bound $\wt{\Bd}^\gp_1$; 
	Bottom-right: Absolute recovery error for $y_2$ and error bound $\wt{\Bd}^\gp_2$.}
	\label{fig:two_chirp_signal2}
\end{figure}

We choose $\mu=1$, and $\gs(b)$ to be $\gs_2(b)$ defined by \eqref{def_gs2}.  
We set $\tau_0=1/20, \wt \ep_1=0.01$, $\wt \ep_3=\frac 12(\phi'_2(b)-\phi'_1(b)$. In addition, we use $u_k$ and $l_k$ given by \eqref{def_ulk}.  
We show the recovered  $y_1(t), y_2(t)$ in the middle row of Fig.\ref{fig:two_chirp_signal2}. 
The absolute recovery errors for $y_1$ and $y_2$ and the error bounds  $\wt{\rm Bd}^\gp_1$ and $\wt{\rm Bd}^\gp_2$  are provided in the bottom row of Fig.\ref{fig:two_chirp_signal2}. From  Fig.\ref{fig:two_chirp_signal2}, we know the recovery errors are small except near boundary points $t=0$ and $t=1$. 
\end{example}

\section*{Appendices}

\subsection*{Appendix A: Proof of Theorem \ref{theo:main_1st}}

In this appendix, we present the proof of Theorem \ref{theo:main_1st}.

\bigskip

{\bf Proof  of  Theorem \ref{theo:main_1st} Part (a)}.  Assume $(a, b)\not \in \cup _{k=1}^K Z_k$. Then for any $k$,  by the definition of $Z_k$ in \eqref{def_Zk}, we have
$|\wh g\big(\gs(b)(\mu -a \phi_k'(b))\big)|\le \tau_0$. Thus, by \eqref{CWT_approx_1st} and \eqref{rem0_est}, we have
\begin{eqnarray*}
|\wt W_x(a, b)|\hskip -0.6cm &&\le \sum_{k=1}^K |x_k(b) \wh g\big(\gs(b)(\mu -a \phi_k'(b))\big)| + |\rem_0|
\\
&&\le a \gs(b) \lambda_0(a, b)+ \tau _0 \sum_{k=1}^K A_k(b)  \\
&& \le  a_2(b) \gs(b) \Lambda_1(b)+ \tau _0 \sum_{k=1}^K A_k(b)  \le \wt \ep_1,
\end{eqnarray*}
a contradiction to the assumption $|\wt W_x(a, b)|>\wt \ep_1$. Therefore, $(a, b)\in Z_\ell$ for some $\ell$. Since $Z_k, 1\le k\le K$ are disjoint, this $\ell$ is unique. Hence, the statement in (a) holds.
\hfill $\blacksquare$

\bigskip

{\bf Proof  of  Theorem \ref{theo:main_1st} Part (b)}.   
By \eqref{CWT_approx_1st} with $g$ replaced by $g'$,
\begin{eqnarray*}
&&\wt W_x^{g'}(a, b)=\sum_{\ell=1}^K \int_\R x_\ell(b)e^{i2\pi  \phi_\ell'(b) a t}\frac 1{\gs(b)}g'\Big(\frac t{\gs(b)}\Big) e^{-i2\pi \mu t}dt +\rem_0'\\
&&=\sum_{\ell=1}^K x_\ell(b)\wh{(g')}\big(\gs(b)(\mu -a \phi_\ell'(b))\big)
 +\rem_0'\\
&&=i2\pi \gs(b)\sum_{\ell=1}^K x_\ell(b)(\mu-a \phi_\ell' (t))\wh g\big(\gs(b)(\mu -a \phi_\ell'(b))\big)
 +\rem_0'.
\end{eqnarray*}
\clearpage 
\n This and \eqref{result_partial_tV} imply that
\begin{eqnarray*}
&& \big( \go_x^{\rm adp, c}(a, b)-\phi_k'(b)\big)i2\pi \wt W_x(a,b)\\
&&=\pd_b \wt W_x(a,b)+\frac {\gs'(b)}{\gs(b)}\big(\wt W_x(a,b)+\wt W_x^{g_3}(a,b)\big)
-i2\pi \phi_k'(b)\wt W_x(a,b)\\
&&=\frac{i2\pi \mu}a  \wt W_x(a,b)-\frac 1{a\gs(b)}\wt W_x^{g'}(a,b)-i2\pi \phi_k'(b)\wt W_x(a,b)\\
&&=i2\pi \Big(\frac \mu a -\phi_k'(b)\Big)\Big (\sum_{\ell=1}^K x_\ell(b)
\wh g\big(\gs(b)(\mu -a \phi_\ell'(b))\big)+\rem_0\Big)\\
&&\qquad -\frac 1{a\gs(b)}\Big(
i2\pi \gs(b)\sum_{\ell=1}^K x_\ell(b)(\mu-a \phi_\ell' (b))\wh g\big(\gs(b)(\mu -a \phi_\ell'(b))\big)
 +\rem_0'
\Big)\\
\end{eqnarray*}
\begin{eqnarray*}&&=i2\pi \Big(\frac \mu a -\phi_k'(b)\Big)\rem_0-\frac{\rem_0'}{a\gs(b)}+
i2\pi \sum_{\ell\not=k} x_\ell(b)(\phi_\ell' (b)-\phi_k' (b))\wh g\big(\gs(b)(\mu -a \phi_\ell'(b))\big)\\
&&=\Rem_1.
\end{eqnarray*}
This shows \eqref{transformation_approx}.

When $(a, b)\in Z_k$, we have $|\frac \mu a -\phi_k'(b)|<\frac\ga{a\gs(b)}$. Thus
\begin{eqnarray*}
&&|\Rem_1|\le 2\pi \ga  \frac {|\rem_0|}{a\gs(b)}  +
\frac {|\rem_0\rq{}|}{a\gs(b)}+ 2\pi \sum_{\ell\not=k} A_\ell(b)|\phi_\ell' (b)-\phi_k' (b)| \;
|\wh g\big(\gs(b)(\mu -a \phi_\ell'(b))\big)|\\
&&\le  2\pi \ga \Lambda_k(b)+\wt \Lambda_k(b)+
2\pi \sum_{\ell\not=k} A_\ell(b)|\phi_\ell' (b)-\phi_k' (b)| \; 
\; \big|\wh g\big(\rho_{\ell, k}(b)\big)\big|\\
&& = 2\pi \wt \ep_1 \bd_k,
\end{eqnarray*}
where the second inequality follows from \eqref{def_Lam0}, \eqref{def_tLam0} and \eqref{rho_ineq}.
Hence,  with the assumptions
$|\wt W_x(a, b)|>\wt \ep_1$,
we have
\begin{eqnarray*}
&&|\go_x^{\rm adp}(a, b)-\phi_k'(b)|\le |\go_x^{\rm adp, c}(a, b)-\phi_k'(b)|\\
&&=\Big| \frac{\Rem_1}{i2\pi\wt W_x(a, b)}
\Big|< \frac{|\Rem_1|}{2\pi \wt \ep_1} \le \bd_k.
\end{eqnarray*}
This proves \eqref{transformation_est}.
\hfill $\blacksquare$

\bigskip

{\bf Proof  of  Theorem \ref{theo:main_1st} Part (c)}.
Following similar discussions in \cite{Daub_Lu_Wu11}, one can obtain that
\begin{equation}
\label{WSST_CWT_relation1_1st}
 \lim_{\gl\to 0}\int_{|\xi-\phi_k'(b)|<\wt \ep_3} T_{x, \wt \ep_1}^{\rm adp, \gl}(\xi, b)d\xi =
 \int_{X_b }  \wt W_x(a, b)
 \frac{da}a ,
\end{equation}
where
$$
X_b :=\big\{a>0:\; |\wt W_x(a, b)|>\wt \ep_1 \;
\hbox{and $\big|\phi_k'(b)-\go_x^{\rm adp}(a, b)\big|<\wt \ep_3$}\big \}.
$$

Next we show that $X_b$ is the set $Y_b$ defined by
$$
Y_b:=\big\{a>0:\; |\wt W_x(a, b)|>\wt \ep_1 \; \hbox{and $(a, b)\in Z_k$}\big \}.
$$
Indeed, by Theorem \ref{theo:main_1st} Part (b), if $a \in Y_b $, then $\big|\phi_k'(b)-\go_x^{\rm adp}(a, b)\big|< \bd_k\le \wt \ep_3$. Thus $a \in X_b $. Hence $Y_b \subseteq X_b $.  On the other hand, suppose $a \in X_b $. Since $|\wt W_x(a, b)|>\wt \ep_1$, by Theorem \ref{theo:main_1st} Part (a), $(a, b)\in Z_\ell$ for an $\ell$ in $\{1, 2, \cdots, K\}$. If $\ell\not =k$, then by Theorem \ref{theo:main_1st} Part (b),
\begin{eqnarray*}
\big|\phi_k'(b)-\go_x^{\rm adp}(a, b)\big|\hskip -0.6cm &&\ge |\phi_k'(b) -\phi'_\ell(b)|-\big|\phi'_\ell(b)-\go_x^{\rm adp}(a, b)\big|\\
&& > \min\{\phi_k'(b) -\phi'_{k-1}(b), \phi_{k+1}'(b) -\phi'_k(b)\}-\bd_\ell\\
&&  \ge \min\{\phi_k'(b) -\phi'_{k-1}(b), \phi_{k+1}'(b) -\phi'_k(b)\}-\wt \ep_3\ge \wt \ep_3.
\end{eqnarray*}
since $\max\limits_{1\le \ell \le K} \{\bd_\ell\}\le \wt \ep_3\le \frac 12 \min\{\phi_k'(b) -\phi'_{k-1}(b), \phi_{k+1}'(b) -\phi'_k(b)\}$.
This contradicts to the assumption $\big|\phi_k'(b)-\go_x^{\rm adp}(a, b)\big|<\wt \ep_3$ since $a \in X_b $. Hence $\ell=k$ and $a\in Y_b $. Thus we get $X_b =Y_b $. This , together with  \eqref{WSST_CWT_relation1_1st}, leads to
\begin{equation}
\label{WSST_CWT_relation2_1st}
 \lim_{\gl\to 0}\int_{|\xi-\phi_k'(b)|<\wt \ep_3} T_{x, \wt \ep_1}^{\rm adp,\gl}(\xi, b)d\xi =
 \int_{\{|\wt W_x(a, b)|>\wt \ep_1\}\cap \{a: (a, b)\in Z_k\}}  \wt W_x(a, b)
 \frac{da}a.
\end{equation}

To prove the estimate \eqref{reconstr_1st}, we consider
\begin{eqnarray*}
&&\Big| \int_{\{|\wt W_x(a, b)|>\wt \ep_1\}\cap \{a: (a, b)\in Z_k\}}  \wt W_x(a, b) \frac{da}a
- c^\ga_\psi(b)x_k(b)\Big|\\
&&= \Big| \int_{\{a: (a, b)\in Z_k\}}  \wt W_x(a, b) \frac{da}a
- \int_{\{|\wt W_x(a, b)|\le\wt \ep_1\}\cap \{a: (a, b)\in Z_k\}}  \wt W_x(a, b) \frac{da}a
- c^\ga_\psi(b)x_k(b)\Big|\\
&&\le  \int_{\{a: (a, b)\in Z_k\}}   \wt \ep_1\frac{da}a
+\Big| \int_{\{a: (a, b)\in Z_k\}}  \big(\sum_{\ell=1}^K x_\ell(b)
\wh g\big(\gs(b)(\mu -a \phi_\ell'(b))\big)
+\rem_0 \big)\frac{da}a - c^\ga_\psi(b)x_k(b)\Big| \\
&& \le \wt \ep_1 \int_{\frac{\mu-\ga/\gs(b)}{\phi_k\rq{}(b)}}^{\frac{\mu+\ga/\gs(b)}{\phi_k\rq{}(b)}}\frac{da}a
+\int_{Z_k} \left |\rem_0\right | \frac {da}a +
\Big| \int_{|\mu -a \phi_k'(b)|<\frac \ga{\gs(b)}}   x_k(b) \wh g\big(\gs(b)(\mu-a \phi_k'(b))\big) \frac{da}a - c^\ga_\psi(b) x_k(b)\Big| \\
&&\qquad + \sum_{\ell\not=k }A_\ell(b) \Big|\int_{|\mu - a \phi_k'(b)|<\frac \ga{\gs(b)}}\wh g\big(\gs(b)(\mu -a \phi_\ell'(b))\big)  \frac{da}a\Big| \\
&&\le \wt \ep_1  \ln \frac {\mu \gs(b)+\ga}{\mu \gs(b)-\ga} 
+  \int_{\frac{\mu-\ga/\gs(b)}{\phi_k\rq{}(b)}}^{\frac{\mu+\ga/\gs(b)}{\phi_k\rq{}(b)}}a \gs(b) \Lambda_k(b) \frac{da}a
+\Big|{x_k(b)} \int_{\mu-\ga/\gs(b)}^{\mu+\ga/\gs(b)}
\wh g\big(\gs(b)(\mu-\xi )\big) \frac{d\xi }\xi
- c^\ga_\psi(b)x_k(b)\Big| \\
&&\qquad +\sum_{\ell\not=k }A_\ell(b)
 \Big|\int_{\mu-\ga/ \gs(b)}^{\mu+\ga/\gs(b)}
 \wh g\big(\gs(b)(\mu-\frac{ \phi_\ell'(b)}{\phi_k'(b)}\xi )\big) \frac{d\xi }\xi  \Big|\\
&&= \wt \ep_1 \ln \frac {\mu \gs(b)+\ga}{\mu \gs(b)-\ga}  + \frac{2\ga}{\phi_k'(b)}
\Lambda_k(b)  +\sum_{\ell\not=k }A_\ell(b)  m_{\ell, k}(b)=| c_\psi^\ga(b)| \; \wt \bd_k.
\end{eqnarray*}
This estimate and \eqref{WSST_CWT_relation2_1st} imply
that \eqref{reconstr_1st} holds.
This completes the proof of Theorem \ref{theo:main_1st} Part (c).
\hfill $\blacksquare$

\subsection*{Appendix B: Proofs of Theorems \ref{theo:main}-\ref{theo:main3}}
In this appendix, we provide the proof of  Theorems \ref{theo:main} and \ref{theo:main3}.

{\bf Proof of  Theorem \ref{theo:main} Part (a)}.  Assume $(a, b)\not \in \cup _{k=1}^K O_k$. Then for any $k$,   
by \eqref{CWT_approx}, \eqref{err0_est}  and \eqref{def_Ok2_ineq},  we have
\begin{eqnarray*}
|\wt W_x(a, b)|\hskip -0.6cm &&\le |\err_0|+\sum_{k=1}^K |x_k(b) G_k\big(\gs(b)(\mu-a \phi_k'(b) )\big)|
\\
&&\le  a\gs(b) \Pi_0(a, b)+\tau _0 \sum_{k=1}^K A_k(b)\\
&&\le  a_2(b)\gs(b) \Pi_0(a_2(b), b)+\tau _0 \sum_{k=1}^K A_k(b)\le \wt \vep_1,
\end{eqnarray*}
a contradiction to the assumption $|\wt W_x(a, b)|>\wt \vep_1$. Thus $(a, b)\in O_\ell$ for some $\ell$. Since $O_k, 1\le k\le K$ are not overlapping, this $\ell$ is unique. This completes the proof of the statement in (a).
\hfill $\blacksquare$

\bigskip

{\bf Proof  of  Theorem \ref{theo:main} Part ${\rm (b)}$}.  Plugging  $\pd_b \wt W_x(a, b)$ in \eqref{result_lem1} to $\go_{x}^{\rm 2adp, c}$ in \eqref{def_transformation_2nd_complex}, we have
\begin{eqnarray*}
&&\go_{x}^{\rm 2adp, c}=\frac {{\partial}_b \wt W_x(a, b)}{i2\pi \wt W_x(a, b)}+\frac{\gs'(b)}{i2\pi \gs(b)}
- a\frac{\wt W^{g_1}_x(a, b)}{i2\pi \wt W_x(a, b)} R_0(a, b)
+ \frac {\gs'(b)}{\gs(b)}\frac {\wt W^{g_3}_x(a, b)}{i2\pi \wt W_x(a, b)}\\
&&=\frac 1{i2\pi \wt W_x(a, b)}\Big\{\Big(i2\pi \phi_k'(b)-\frac {\gs'(b)}{\gs(b)}\Big)\wt W_x(a, b)+
i2\pi \phi''_k(b)a\gs(b)\wt W^{g_1}_x(a, b)- \frac {\gs'(b)}{\gs(b)}\wt W^{g_3}_x(a, b)+\Err_1\Big\}
\\
&&\qquad + \frac{\gs'(b)}{i2\pi \gs(b)} - a\frac{\wt W^{g_1}_x(a, b)}{i2\pi \wt W_x(a, b)} R_0(a, b)
+ \frac {\gs'(b)}{\gs(b)}\frac {\wt W^{g_3}_x(a, b)}{i2\pi \wt W_x(a, b)}
\\
&&=\phi_k'(b)+ \phi''_k(b)a\gs(b)\frac{\wt W^{g_1}_x(a, b)}{\wt W_x(a, b)}+\frac{\Err_1}{i2\pi \wt W_x(a, b)} - a\frac{\wt W^{g_1}_x(a, b)}{i2\pi\wt W_x(a, b)} R_0(a, b)\\
&&=\phi_k'(b)+ \phi''_k(b)a\gs(b)\frac{\wt W^{g_1}_x(a, b)}{\wt W_x(a, b)}+\frac{\Err_1}{i2\pi \wt W_x(a, b)} - a\frac{\wt W^{g_1}_x(a, b)}{i2\pi\wt W_x(a, b)} \big(i2\pi \gs(b)
\phi''_k(b)+\Err_3\big)\\
&&=\phi_k'(b)+\frac{\Err_1}{i2\pi \wt W_x(a, b)}- a\frac{\wt W^{g_1}_x(a, b)\Err_3}{i2\pi\wt W_x(a, b)} \\
&&=\phi_k'(b)+\Err_4,
\end{eqnarray*}
where \eqref{result_lem3} has been used above. Thus \eqref{transformation_approx_2nd} holds. 

To prove \eqref{IF_est_Bd1}, observe that
$$
\Err_3=\frac { \frac{\Err_2}{\wt W_x(a, b)}  -\frac{\pd _a \wt W_x(a, b) \; \Err_1}{\wt W_x(a, b)^2} }
{\pd _a \Big( \frac {a \wt W^{g_1}_x(a, b)}{\wt W_x(a, b)}\Big)}.
$$
Thus for $(a, b)\in O_k$ and $|\wt W_x(a, b)|\ge \wt \vep_1$ and
$\big|\pd_a\Big( \frac {a \wt W^{g_1}_x(a, b)}{\wt W_x(a, b)}\Big)\big|\ge \wt \vep_2$, we have
$$
|\Err_3|\le \frac 1{\wt \vep_2} \Big(\frac {|\Err_2|}{\wt \vep_1}+  \frac{|\pd _a \wt W_x(a, b) \; \Err_1|}{\wt \vep_1^2} \Big)=\frac 1{\wt \vep_1^2 \wt \vep_2} (|\Err_2| \wt \vep_1+  |\pd _a \wt W_x(a, b)|  \; |\Err_1|).
$$
Hence
\begin{eqnarray}
\nonumber
|\Err_4|
\hskip -0.6cm &&=\Big| \frac{\Err_1}{i2\pi \wt W_x(a, b)}- a\frac{\wt W^{g_1}_x(a, b)\Err_3}{i2\pi\wt W_x(a, b)}
\Big|\\
 \label{est_Res4} &&< \frac{|\Err_1|}{2\pi \wt \vep_1}+ \frac1{2\pi \wt \vep_1^3 \wt \vep_2}
|a \wt W^{g_1}_x(a, b)| \big(|\Err_2| \wt \vep_1+ |\pd _a \wt W_x(a, b)|\;  |\Err_1| \big)\\
&&\le \Bd_k.
\nonumber
\end{eqnarray}
This proves \eqref{IF_est_Bd1}. 
\hfill $\blacksquare$

\bigskip

{\bf Proof  of  Theorem \ref{theo:main} Part (c)}. First we have the following result which can be derived as that on p.254 in \cite{Daub_Lu_Wu11}:
\begin{equation}
\label{WSST_CWT_relation1}
 \lim_{\gl\to 0}\int_{|\xi-\phi_k'(b)|<\wt \vep_3} T_{x, \wt \vep_1, \wt \vep_2}^{\rm 2adp, \gl}(\xi, b)d\xi =
 \int_{Z_b }  \wt W_x(a, b)
 \frac{da}a ,
\end{equation}
where
\begin{equation*}
Z_b:=\big\{a:\; |\wt W_x(a, b)|>\wt \vep_1,  \;  \big| \partial _a\big(a{\wt W^{g_1}_x(a, b)}/{\wt W_x(a, b)}\big)\big| > \wt \vep_2 \;
\hbox{and $\big|\phi_k'(b)-\go_{x, \wt \vep_2}^{\rm 2adp}(a, b)\big|<\wt \vep_3$}\big \}.
\end{equation*}

Let $V_b$ be the set defined by \eqref{def_Yt}. Next we show that $V_b =Z_b $. First we have that if $a\in V_b $, then by Theorem \ref{theo:main} Part (b),  $\big|\phi_k'(b)-\go_{x, \wt \vep_2}^{\rm 2adp}(a, b)\big|<\Bd_k\le \wt \vep_3$. Thus $a \in Z_b $. Hence we have  $V_b \subseteq Z_b $.

On the other hand, suppose $a \in Z_b $. Since $|\wt W_x(a, b)|>\wt \vep_1$, by Theorem \ref{theo:main} Part (a), $(a, b)\in O_\ell$ for an $\ell$ in $\{1, 2, \cdots, K\}$. If $\ell\not =k$, then
\begin{eqnarray*}
\big|\phi_k'(b)-\go_{x, \wt \vep_2}^{\rm 2adp}(a, b)\big|\hskip -0.6cm &&\ge |\phi_k'(b) -\phi'_\ell(b)|-\big|\phi'_\ell(b)-\go_{x, \wt \vep_2}^{\rm 2adp}(a, b)\big|\\
&&> L_k(b)-\Bd_\ell\ge L_k(b)-\wt \vep_3\ge \wt \vep_3,
\end{eqnarray*}
and this contradicts to the assumption $a \in Z_b $ with $\big|\phi_k'(b)-\go_{x, \wt \vep_2}^{\rm 2adp}(a, b)\big|<\wt \vep_3$, where we have used the fact $|\phi_k'(b) -\phi'_\ell(b)|\ge L_k(b)$ and
$\big|\phi'_\ell(b)-\go_{x, \wt \vep_2}^{\rm 2adp}(a, b)\big|<\Bd_k\le \wt \vep_3$ by Theorem \ref{theo:main} Part (b).
Hence $\ell=k$ and $a \in V_b $. Therefore $V_b =Z_b$.

\bigskip

The facts $Z_b =V_b $ and $V_b \cap U_b=\emptyset$, together with \eqref{WSST_CWT_relation1}, imply that
\begin{eqnarray}
 \nonumber &&\lim_{\gl\to 0}\int_{|\xi-\phi_k'(b)|<\wt \vep_3} T_{x, \wt \vep_1, \wt \vep_2}^{\rm 2adp, \gl}(\xi, b)d\xi =
 \int_{V_b }\wt W_x(a, b) \frac{da}a  =\int_{V_b \cup U_b}\wt W_x(a, b) \frac{da}a - \int_{U_b}\wt W_x(a, b) \frac{da}a \\
\label{WSST_CWT_relation2} && = \int_{\{|\wt W_x(a, b)|>\wt \vep_1\}\cap \{a: (a, b)\in O_k\}}  \wt W_x(a, b)
 \frac{da}a - \int_{U_b}\wt W_x(a, b) \frac{da}a.
\end{eqnarray}

Furthermore,
\begin{eqnarray*}
&&\Big|\int_{\{|\wt W_x(a, b)|>\wt \ep_1\}\cap \{a: (a, b)\in O_k\}}  \wt W_x(a, b) \frac{da}a
- c^k_\psi(b)x_k(b)\Big|\\
&&= \Big| \int_{\{a: (a, b)\in O_k\}}  \wt W_x(a, b) \frac{da}a
- \int_{\{|\wt W_x(a, b)|\le\wt \ep_1\}\cap \{a: (a, b)\in O_k\}}  \wt W_x(a, b) \frac{da}a
- c^k_\psi(b)x_k(b)\Big|\\
&&\le  \int_{\{a: (a, b)\in O_k\}}   \wt \ep_1\frac{da}a
+\Big| \int_{\{a: (a, b)\in O_k\}}  \big(\sum_{\ell=1}^K x_\ell(b)
G_k\big(\gs(b)(\mu -a \phi_\ell'(b))\big)
+\err_0 \big)\frac{da}a - c^k_\psi(b)x_k(b)\Big| \\
&& \le \wt \ep_1 \int_{l_k}^{u_k} \frac{da}a
+\int_{l_k}^{u_k} |\err_0| \frac {da}a +
\Big| \int_{l_k}^{u_k}  x_k(b) G_k\big(\gs(b)(\mu-a \phi_k'(b))\big) \frac{da}a - c^k_\psi(b) x_k(b)\Big| \\
&&\qquad + \sum_{\ell\not=k }A_\ell(b) \Big|\int_{l_k}^{u_k} G_k\big(\gs(b)(\mu -a \phi_\ell'(b))\big)  \frac{da}a\Big| \\
&&\le \wt \ep_1 \ln \frac {u_k(b)}{l_k(b)}
+  \int_{l_k}^{u_k} a \gs(b) \Pi_0(a, b) \frac{da}a
+\big|{x_k(b)}c^k_\psi(b)
- c^k_\psi(b)x_k(b)\big| + \sum_{\ell\not=k }A_\ell(b)  M_{\ell, k}(b)\\
&&= \wt \ep_1 \ln \frac {u_k(b)}{l_k(b)}
 +\gs(b) K \vep_1  I_1(u_k-l_k) +\frac \pi 9 \vep_3 I_3 (u_k-l_k)^3 \gs^3(b) \sum_{j=1}^K A_j(b)
  +\sum_{\ell\not=k }A_\ell(b)  M_{\ell, k}(b)\\
  && =\wt \Bd'_k.
\end{eqnarray*}
Hence, we have
\begin{equation}
\label{est_Bd3p}
\Big|\frac 1{c^k_\psi(b)} \int_{\{|\wt W_x(a, b)|>\wt \vep_1\}\cap \{a: (a, b)\in O_k\}}  \wt W_x(a, b) \frac{da}a
-x_k(b)\Big|\le \frac 1{|c^k_\psi(b)|}\wt \Bd'_k.
\end{equation}

In addition, 
\begin{eqnarray*}
&&\Big|\int_{U_b}  \wt W_x(a, b) \frac{da}a\Big|=
\Big| \int_{U_b}  \big(\sum_{\ell=1}^K x_\ell(b)
G_k(\gs(b)(\mu - a\phi'_k(b))+\err_0 \big)\frac{da}a \Big| \\
&& \le    \int_{\{a: (a, b)\in O_k\}}  |\err_0| \frac{da}a +\frac{A_k(b)}{l_k(b)} \sup_{a \in U_b}
|G_k(\gs(b)(\mu -a\phi'_k(b))| \; |U_b|\\
&&\qquad
+\sum_{\ell\not=k }A_\ell(b) \int_{\{a: (a, b)\in O_k\}} |G_k(\gs(b)(\mu - a\phi'_k(b))| \frac{da}a
\\
&&
\le \gs(b) K \vep_1  I_1(u_k-l_k) +\frac \pi 9 \vep_3 I_3 (u_k-l_k)^3 \gs^3(b) \sum_{j=1}^K A_j(b)
+\frac{A_k(b)}{l_k(b)} \|g\|_1 \; |U_b|+\sum_{\ell\not=k }A_\ell(b) M_{\ell, k}(b)
\\
&&= \wt \Bd''_k,
\end{eqnarray*}
where we have used the fact
$$
\sup_\xi |G_k(\xi)|\le \int_{\R} |e^{i\pi\gs^2(b) \phi''_k(b) a^2 t^2}g(t) e^{-i2\pi \xi t}|dt=\|g\|_1.
$$
The above estimates, together with \eqref{WSST_CWT_relation2}, leads to \eqref{reconstr}. This completes the proof of Theorem \ref{theo:main} Part (c).
\hfill $\blacksquare$

\bigskip

Theorem \ref{theo:main3} Part ${\rm (b_1)}$ follows immediately from \eqref{est_Res4}.

\bigskip
{\bf Proof  of  Theorem \ref{theo:main3} Part ${\rm (b_2)}$}. By \eqref{result_lem1} in Lemma 1, we have
\begin{eqnarray*}
&&\go_{x}^{\rm adp, c}=\frac {{\partial}_b \wt W_x(a, b)}{i2\pi \wt W_x(a, b)}+
\frac{\gs'(b)}{i2\pi \gs(b)}+ \frac {\gs'(b)}{\gs(b)}\frac {\wt W^{g_3}_x(a, b)}{i2\pi \wt W_x(a, b)}\\
&&=\frac 1{i2\pi \wt W_x(a, b)}\Big\{\big(i2\pi \phi_k'(b)-\frac {\gs'(b)}{\gs(b)}\big)\wt W_x(a, b)+
i2\pi \phi''_k(b)a\gs(b)\wt W^{g_1}_x(a, b)- \frac {\gs'(b)}{\gs(b)}\wt W^{g_3}_x(a, b)+\Err_1\Big\}
\\
&&\qquad + \frac{\gs'(b)}{i2\pi \gs(b)}+ \frac {\gs'(b)}{\gs(b)}\frac {\wt W^{g_3}_x(a, b)}{i2\pi \wt W_x(a, b)}\\
&&=\phi_k'(b)+ \phi''_k(b)a\gs(b)\frac{\wt W^{g_1}_x(a, b)}{\wt W_x(a, b)}+\frac{\Err_1}{i2\pi \wt W_x(a, b)}.
\end{eqnarray*}
This shows \eqref{transformation_approx_1st_chirp}. \eqref{transformation_est_1st_chirp3} follows from  \eqref{transformation_approx_1st_chirp} and the assumption $|\wt W_x(a, b)|> \wt \vep_1$. \hfill $\blacksquare$

\bigskip

{\bf Proof  of  Theorem \ref{theo:main3} Part (c)}. First we have the following result which can be derived as that on p.254 in \cite{Daub_Lu_Wu11}:
\begin{equation}
\label{WSST_CWT_relation1_3}
 \lim_{\gl\to 0}\int_{|\xi-\phi_k'(b)|<\wt \vep_3} S_{x, \wt \vep_1, \wt \vep_2}^{\rm 2adp, \gl}(\xi, b)d\xi =
 \int_{\wt X_b }  \wt W_x(a, b)
 \frac{da}a ,
\end{equation}
where
\begin{equation*}
\wt X_b:=\big\{a>0:\; |\wt W_x(a, b)|>\wt \vep_1 \;
\hbox{and $\big|\phi_k'(b)-\go_{x, \wt \vep_2}^{\rm 2adp}(a, b)\big|<\wt \vep_3$}\big \}.
\end{equation*}

Let
\begin{equation*}
\wt Y_b:=\big\{a>0:\; |\wt W_x(a, b)|>\wt \vep_1 \; \hbox{and $(a, b)\in O_k$}\big \}.
\end{equation*}
Next we show that $\wt X_b=\wt Y_b$. By Theorem \ref{theo:main3} Part (b${}_1$)(b${}_2$), if $a \in \wt Y_b $, then $\big|\phi_k'(b)-\go_{x, \wt \vep_2}^{\rm 2adp}(a, b)\big|<\wt \vep_3$ since $\Bd_1', \Bd_2\rq{}\le \wt \vep_3$. Thus $a \in \wt X_b $. Hence $\wt Y_b \subseteq \wt X_b $.

On the other hand, suppose $a \in \wt X_b $. Since $|\wt W_x(a, b)|>\wt \vep_1$, by Theorem \ref{theo:main} Part (a), $(a, b)\in O_\ell$ for an $\ell$ in $\{1, 2, \cdots, K\}$. If $\ell\not =k$, then
\begin{eqnarray*}
\big|\phi_k'(b)-\go_{x, \wt \vep_2}^{\rm 2adp}(a, b)\big|\hskip -0.6cm &&\ge |\phi_k'(b) -\phi'_\ell(b)|-\big|\phi'_\ell(b)-\go_{x, \wt \vep_2}^{\rm 2adp}(a, b)\big|\\
&&> L_k(b)-\max\{\Bd_1', \Bd_2\rq{}\}\ge  L_k(b)-\wt \vep_3\ge \wt \vep_3,
\end{eqnarray*}
and this contradicts to the assumption $a \in \wt X_b $ with $\big|\phi_k'(b)-\go_{x, \wt \vep_2}^{\rm 2adp}(a, b)\big|<\wt \vep_3$, where we have used the fact $|\phi_k'(b) -\phi'_\ell(b)|\ge L_k(b)$ and
$\big|\phi'_\ell(b)-\go_{x, \wt \vep_2}^{\rm 2adp}(a, b)\big|<\max(\Bd_1', \Bd_2\rq{})\le \wt \vep_3$ by Theorem \ref{theo:main3} Part (b${}_1$)(b${}_2$).
Hence $\ell=k$ and $a \in \wt Y_b $. Thus we know $\wt X_b =\wt Y_b $. This and \eqref{WSST_CWT_relation1_3} imply
\begin{equation}
\label{WSST_CWT_relation2_3}
 \lim_{\gl\to 0}\int_{|\xi-\phi_k'(b)|<\wt \vep_3} S_{x, \wt \vep_1, \wt \vep_2}^{\rm 2adp, \gl}(\xi, b)d\xi =
 \int_{\{|\wt W_x(a, b)|>\wt \vep_1\}\cap \{a: (a, b)\in O_k\}}  \wt W_x(a, b)
 \frac{da}a.
\end{equation}

The estimate \eqref{est_Bd3p}, together with \eqref{WSST_CWT_relation2_3}, leads to \eqref{reconstr3}.
This completes the proof of Theorem \ref{theo:main3} Part (c).
\hfill $\blacksquare$

\subsection*{Appendix C: Proofs of Lemmas \ref{lem:lem0}-\ref{lem:lem3}}

In this appendix, we provide the proof of Lemmas \ref{lem:lem0}-\ref{lem:lem3}. For simplicity of presentation,  we drop $x, a, b$ in  $\wt W_x(a, b), \wt W^{g\rq{}}_x(a, b), \wt W^{g_j}_x(a, b)$ below.

\bigskip 

{\bf Proof  of  Lemma \ref{lem:lem0}}. \quad
By \eqref{def_CWT_para0},  we have 
\begin{eqnarray*}
&&\pd_b\wt W=\int_{-\infty}^\infty x(t)\pd_b\Big\{ \frac 1{a\gs(b)} g\Big(\frac{ t-b}{a\gs(b)}\Big)e^{-i2\pi \mu \frac{t-b}a} \Big\}dt\\
&&=\int_{-\infty}^\infty x(t)\Big\{ -\frac {\gs\rq{}(b)}{a\gs^2(b)} g\Big(\frac{ t-b}{a\gs(b)}\Big)
+\frac 1{a\gs(b)}g\rq{}\Big(\frac{ t-b}{a\gs(b)}\Big)\Big(-\frac 1{a\gs(b)}- \frac {\gs\rq{}(b)}{\gs^2(b)}\frac{t-b}a\Big)
\Big\}e^{-i2\pi \mu \frac{t-b}a}dt  \\
&& \qquad + \int_{-\infty}^\infty x(t) \frac 1{a\gs(b)} g\Big(\frac{ t-b}{a\gs(b)}\Big)e^{-i2\pi \mu \frac{t-b}a} \frac {i2\pi\mu}a dt\\
&&=-\frac{\gs\rq{}(b)}{\gs(b)} \wt W -\frac 1{a\gs(b)}\wt W^{g\rq{}}-\frac{\gs\rq{}(b)}{\gs(b)} \wt W^{g_3}+\frac {i2\pi\mu}a \wt W, 
\end{eqnarray*}
which is the right-hand side of \eqref{result_partial_tV}.  Thus \eqref{result_partial_tV} holds. 
\hfill $\blacksquare$

\bigskip

{\bf Proof  of  Lemma \ref{lem:lem1}}. \quad
By \eqref{CWT_approx0} with $g$ replaced by $g'$,
\begin{eqnarray*}
&&\wt W^{g'}=\sum_{\ell=1}^K \int_\R x_\ell(b)e^{i2\pi (\phi_\ell'(b) at +\frac12\phi''_\ell(b) a^2t^2)}\frac 1{\gs(b)}g'\Big(\frac t{\gs(b)}\Big) e^{-i2\pi \mu t}dt +\err_0'\\
&&=\sum_{\ell=1}^K \int_\R x_\ell(b)e^{-i2\pi (\mu -a \phi_\ell'(b)) t +i\pi \phi''_\ell(b) a^2 t^2} \frac {\pd}{\pd t} \Big(g\Big(\frac t{\gs(b)}\Big)\Big) dt +\err_0'\\
&&=-\sum_{\ell=1}^K \int_\R \frac {\pd}{\pd t} \Big( x_\ell(b)e^{-i2\pi (\mu-a \phi_\ell' (b) )t +i\pi \phi''_\ell(b) a^2 t^2}\Big)g\Big(\frac t{\gs(b)}\Big) dt +\err_0'\\
&&=i2\pi \sum_{\ell=1}^K x_\ell(b)(\mu-a \phi_\ell' (b))\int_\R  e^{-i2\pi (\mu-\phi_\ell' (b) )t +i\pi \phi''_\ell(b) a^2 t^2}g\Big(\frac t{\gs(b)}\Big) dt\\
&&\quad  -i2\pi \sum_{\ell=1}^K x_\ell(b)\phi_\ell'' (b)
a^2 \int_\R  e^{-i2\pi (\mu-a \phi_\ell' (b))t +i\pi \phi''_\ell(b) a^2 t^2}t g\Big(\frac t{\gs(b)}\Big) dt+\err_0'
\end{eqnarray*}
\begin{eqnarray*}
&&=i2\pi \gs(b)\sum_{\ell=1}^K x_\ell(b)(\mu-a \phi_\ell' (b))G_{0, \ell}(a, b)
 - i2\pi a^2\gs^2(b)\sum_{\ell=1}^K x_\ell(b)\phi_\ell'' (b)G_{1, \ell}(a, b)  +\err_0'.
\end{eqnarray*}
This and \eqref{result_partial_tV} imply that
\begin{eqnarray*}
&&\pd_b \wt W+\frac {\gs'(b)}{\gs(b)}(\wt W+\wt W^{g_3})
-i2\pi \phi_k'(b)\wt W-i2\pi \phi''_k(b)a\gs(b)\wt W^{g_1}\\
&&=\frac{i2\pi \mu}a \wt W-\frac 1{a\gs(b)}\wt W^{g'}-i2\pi \phi_k'(b)\wt W-i2\pi \phi''_k(b)a\gs(b)\wt W^{g_1}\\
&&=\frac{i2\pi \mu}a \wt W-\frac {i2\pi}{a}
\sum_{\ell=1}^K x_\ell(b)(\mu-a \phi_\ell' (b))G_{0, \ell}(a, b)
 + i2\pi a\gs(b)\sum_{\ell=1}^K x_\ell(b)\phi_\ell'' (b)G_{1, \ell}(a, b)  -\frac{\err_0'}{a\gs(b)}
\\
&&\qquad -i2\pi \phi_k'(b)\wt W-i2\pi \phi''_k(b)a\gs(b)\wt W^{g_1}\\
&&=\frac{i2\pi}a \big(\mu -a\phi_k'(b)\big)\Big (\sum_{\ell=1}^K x_\ell(b)G_{0, \ell}(a, b)+\err_0\Big)\\
&&\qquad -\frac{i2\pi} a \sum_{\ell=1}^K x_\ell(b)(\mu-a \phi_\ell' (b))G_{0, \ell}(a, b)
 + i2\pi a\gs(b)\sum_{\ell=1}^K x_\ell(b)\phi_\ell'' (b)G_{1, \ell}(a, b)  -\frac{\err_0'}{a\gs(b)}\\
 &&\qquad
-i2\pi \phi_k''(b)a \gs(b)\Big (\sum_{\ell=1}^K x_\ell(b)G_{1, \ell}(a, b)+\err_1\Big)
\\
&&=i2\pi \sum_{\ell\not=k} x_\ell(b)(\phi_\ell' (b)-\phi_k' (b))G_{0, \ell}(a, b) +
i2\pi a \gs(b)\sum_{\ell\not= k} x_\ell(b)(\phi_\ell'' (b)-\phi''_k (b))G_{1, \ell}(a, b)\\
&&\qquad +i2\pi \Big(\frac \mu a -\phi_k'(b)\Big)\err_0-\frac{\err_0'}{a \gs(b)}-i2\pi \phi''_k(b)a \gs(b) \; \err_1\\
&&=i2\pi B_k(a, b)+i2\pi a \gs(b) D_k(a, b)+ i2\pi \Big(\frac \mu a -\phi_k'(b)\Big)\err_0-\frac{\err_0'}{a\gs(b)}-i2\pi \phi''_k(b)a\gs(b)\;  \err_1\\
&&=\Err_1.
\end{eqnarray*}
This completes the proof of Lemma \ref{lem:lem1}. \hfill $\blacksquare$

\bigskip

{\bf Proof of Lemma \ref{lem:lem2}}. \quad  \eqref{result_lem2} follows immediately 
from  \eqref{result_lem1} if $\pd_a \Err_1=\Err_2$.  Thus to prove  Lemma \ref{lem:lem2}, it is enough to show $\pd_a \Err_1=\Err_2$. By the definition of $G_{j,k}$ in \eqref{def_Gjk},  
one can easily obtain that for $j\ge 0$,
\begin{eqnarray*}
\partial_a G_{j, k}(a, b)= i2\pi \gs(b) \phi_k'(b) G_{j+1, k}(a, b)
+i2\pi a \gs^2(b)\phi''_k(b)G_{j+2, k}(a, b).
\end{eqnarray*}
By this and direct calculations, one can get
 $\pd_a \Err_{1,1}=\Err_{2,1}$.
So we need merely to show $\pd_a \Err_{1, 2}=\Err_{2,2}$.
To this regard, first we notice that
$$
\pd_a \big(x_{\rm r}(a,b,t)\big)=\frac ta \pd_t \big( x_{\rm r}(a,b,t)\big).
$$
This follows from $\pd_a \big(x(b+a t)\big)=\frac ta \pd_t \big( x(b+ at)\big)$ and $\pd_a \big(x_{\rm m}(a,b,t)\big)=\frac ta \pd_t \big( x_{\rm m}(a,b,t)\big)$. The latter can be verified straightforward by the definition of $x_{\rm m}(a,b,t)$. Thus, we have
\begin{eqnarray*}
\pd_a \err_0 \hskip -0.6cm &&=  \int_\R \pd_a \big( x_{\rm r}(a,b,t)\big) \frac 1{\gs(b)}g\Big(\frac t{\gs(b)}\Big) e^{-i2\pi \mu t}dt\\
&&=  \int_\R \frac ta \pd_t \big( x_{\rm r}(a,b,t)\big) \frac 1{\gs(b)}g\Big(\frac t{\gs(b)}\Big) e^{-i2\pi \mu t}dt\\
&& = \frac 1a (-1)\int_\R  x_{\rm r}(a,b,t) \frac 1{\gs(b)} \pd_t\Big(t g\Big(\frac t{\gs(b)}\Big) e^{-i2\pi \mu t}\Big)dt\\
&& = -\frac 1a \int_\R  x_{\rm r}(a,b,t) \frac 1{\gs(b)} \Big( g\Big(\frac t{\gs(b)}\Big)+\frac t{\gs(b)}g^\gp(\frac t{\gs(b)})
-i2\pi \mu t g\Big(\frac t{\gs(b)}\Big) \Big) e^{-i2\pi \mu t}dt.
\end{eqnarray*}
Therefore,
\begin{eqnarray}
\label{der_res0}
&&\pd_a \err_0
=-\frac 1a\big(\err_0 +\err^\gp_1 -i2\pi \mu \gs(b) \; \err_1\big).
\end{eqnarray}
One can show similarly that
\begin{eqnarray}
\label{der_res1}
&&\pd_a \err_1
=-\frac 1a\big(2\err_1 +\err^\gp_2 -i2\pi \mu \gs(b) \; \err_2\big).
\end{eqnarray}
In addition, from \eqref{der_res0}, we have
\begin{eqnarray}
\label{der_der_res0}
&&\pd_a \err^\gp_0
=-\frac 1a\big(\err^\gp_0 +\err^{\gp\gp}_1 -i2\pi \mu \gs(b) \; \err^\gp_1\big).
\end{eqnarray}
Finally, by \eqref{der_res0}-\eqref{der_der_res0} and
tedious calculations,
one can obtain $\pd_a \Err_{1, 2}=\Err_{2,2}$. This shows $\pd_a \Err_1=\Err_2$ and hence Lemma \ref{lem:lem2} holds.
\hfill $\blacksquare$

\bigskip

{\bf Proof of Lemma \ref{lem:lem3}}. \quad
Note that
$$
R_0(a, b)=\frac1{\wt W \wt W^{g_1} + a \wt W\pd _a
\wt W^{g_1}-a \wt W^{g_1}\pd _a \wt W}\Big(\wt W\pd _a\pd_b\wt W-\pd _a\wt W\pd_b
\wt W+\frac {\gs'(b)}{\gs(b)}(\wt W\pd _a\wt W^{g_3}-\wt W^{g_3}\pd _a\wt W) \Big).
$$
Thus, by \eqref{result_lem1} and \eqref{result_lem2},
\begin{eqnarray*}
&&\big(R_0(a, b)-i2\pi \gs(b)\phi^{\prime\prime}_k(b)\big)
\big(\wt W \wt W^{g_1} + a \wt W\pd _a
\wt W^{g_1}-a \wt W^{g_1}\pd _a \wt W\big)\\
&&=\wt W\pd _a\pd_b\wt W-\pd _a\wt W\pd_b\wt W
+\frac {\gs'(b)}{\gs(b)}(\wt W\pd _a
\wt W^{g_3}-\wt W^{g_3}\pd _a
\wt W)\\
&&\qquad  -i2\pi \gs(b)\phi''_k(b)\big(\wt W \wt W^{g_1} + a \wt W\pd _a
\wt W^{g_1}-a \wt W^{g_1}\pd _a \wt W\big)
\\
&&=\wt W\Big(
\big(i2\pi \phi_k'(b)-\frac {\gs'(b)}{\gs(b)}\big)\pd _a
\wt W+i2\pi \phi''_k(b)\gs(b)(\wt W^{g_1}+a\pd _a \wt W^{g_1})
- \frac {\gs'(b)}{\gs(b)}\pd _a \wt W^{g_3}+\Err_2
\Big)\\
&&\qquad -\pd _a\wt W\Big(
\big(i2\pi \phi_k'(b)-\frac {\gs'(b)}{\gs(b)}\big)\wt W+
i2\pi \phi''_k(b)a \gs(b)\wt W^{g_1}- \frac {\gs'(b)}{\gs(b)}\wt W^{g_3}+\Err_1
\Big)\\
&&\qquad +\frac {\gs'(b)}{\gs(b)}(\wt W\pd _a
\wt W^{g_3}-\wt W^{g_3}\pd _a
\wt W) -i2\pi \gs(b)\phi^{\prime\prime}_k(b)
\big(\wt W \wt W^{g_1} + a \wt W\pd _a
\wt W^{g_1}-a \wt W^{g_1}\pd _a \wt W\big)\\
&&=\wt W\; \Err_2 -\pd _a\wt W\; \Err_1.
\end{eqnarray*}
Therefore,  we have
$$
R_0(a, b)-i2\pi \gs(b)\phi^{\prime\prime}_k(b)=
\frac{\wt W\; \Err_2 -\pd _a\wt W\; \Err_1}
{\wt W \wt W^{g_1} + a \wt W\pd _a
\wt W^{g_1}-a \wt W^{g_1}\pd _a \wt W}
=\Err_3,
$$
as desired.  This completes the proof of Lemma \ref{lem:lem3}. \hfill $\blacksquare$



\end{document}